\documentclass[
amsmath,
prd,nofootinbib,floatfix,11pt,
]{revtex4}

\usepackage{graphicx}
\usepackage{setspace}
\usepackage{bm}
\usepackage[dvips]{color}
\definecolor{Red}{cmyk}{0,1,1,0}
\definecolor{BrickRed}{cmyk}{0,0.89,0.94,0.28}
\definecolor{Blue}{cmyk}{1,1,0,0}
\definecolor{Green}{cmyk}{1,0,1,0}

\newcommand\beq{\begin{eqnarray}}
\newcommand\eeq{\end{eqnarray}}
\def\lsim{\mathrel{\rlap{\lower4pt\hbox{$\sim$}}
    \raise1pt\hbox{$<$}}}                
\def\gsim{\mathrel{\rlap{\lower4pt\hbox{$\sim$}}
    \raise1pt\hbox{$>$}}}            

\allowdisplaybreaks
\newcommand\lnbar{\overline{\ln}}

\newcommand\MSbar{$\overline{\rm{MS}}$ }

\begin{document}

\renewcommand{\theequation}{\arabic{section}.\arabic{equation}}
\renewcommand{\thefigure}{\arabic{section}.\arabic{figure}}
\renewcommand{\thetable}{\arabic{section}.\arabic{table}}

\title{\Large \baselineskip=20pt 
Standard Model parameters in the tadpole-free pure $\overline{\rm{MS}}$ scheme}

\author{Stephen P.~Martin$^{1}$ and David G.~Robertson$^2$}
\affiliation{
\mbox{\it $^1$Department of Physics, Northern Illinois University, DeKalb IL 60115}\\
\mbox{\it $^2$Department of Physics, Otterbein University, Westerville OH 43081}}

\begin{abstract}\normalsize \baselineskip=15pt 
We present an implementation and numerical study of the Standard Model couplings, masses, and vacuum expectation value (VEV), using the pure \MSbar renormalization scheme based on dimensional regularization. Here, the \MSbar Lagrangian parameters are treated as the fundamental inputs, and the VEV is defined as the minimum of the Landau gauge  effective potential, so that tadpole diagrams vanish, resulting in improved convergence of perturbation theory. State-of-the-art calculations relating the \MSbar inputs to on-shell observables are implemented in a consistent way within a public computer code library, {\tt SMDR} (Standard Model in Dimensional Regularization), which can be run interactively or called by other programs. Included here for the first time are the full 2-loop contributions to the Fermi constant within this scheme and studies of the minimization condition for the VEV at 3-loop order with 4-loop QCD effects. We also implement, and study the scale dependence of, all known multi-loop contributions to the physical masses of the Higgs boson, the $W$ and $Z$ bosons, and the top quark, the fine structure constant and weak mixing angle, and the renormalization group equations and threshold matching relations for the gauge couplings, fermion masses, and Yukawa couplings.
\end{abstract}

\maketitle

\vspace{-0.2in}

\tableofcontents

\baselineskip=15.4pt

\newpage

\section{Introduction \label{sec:intro}}
\setcounter{equation}{0}
\setcounter{figure}{0}
\setcounter{table}{0}
\setcounter{footnote}{1}

With the discovery of the Higgs boson, the Standard Model 
is technically complete. This is despite indications that 
it will have to be extended to accommodate dark matter and to solve issues 
such as the hierarchy problem, the strong CP problem, and the cosmological
constant problems. At this writing, 
the LHC continues to strengthen lower bounds on the masses of new particles 
in hypothetical ultraviolet completions such as supersymmetry. It is 
therefore plausible that we should view the Standard Model as a valid, complete 
effective field theory up to the TeV scale and perhaps well beyond, with 
non-renormalizable terms in the Lagrangian correspondingly highly suppressed.
This paper is concerned with the ongoing program of determining, as accurately 
as possible, the relations between the renormalizable 
Lagrangian parameters that define
the theory and the observables and on-shell quantities that are more directly 
connected to experimental results. This is part of a larger 
goal of improving our understanding of
the Standard Model at the level of accuracy required to test it with future experiments.

A convenient method of handling the ultraviolet divergences of 
the Standard Model is provided by dimensional regularization 
\cite{Bollini:1972ui,Ashmore:1972uj,Cicuta:1972jf,tHooft:1972fi,tHooft:1973mm}
followed by renormalization by modified minimal subtraction, $\overline{\mbox{MS}}$
\cite{Bardeen:1978yd,Braaten:1981dv}. 
To describe the effects of electroweak
symmetry breaking induced by the Higgs VEV, there are at least two distinct
ways to proceed. Consider the Higgs potential
\beq
V(\phi) = \Lambda + m^2 H^\dagger H + \lambda (H^\dagger H)^2,
\eeq
where $H$ is the canonically normalized complex Higgs doublet field.
First, one may choose to organize perturbation theory by expanding 
the electrically neutral component of $H$
around a tree-level VEV $v_{\rm tree}/\sqrt{2}$, defined by:
\beq
v_{\rm tree} &\equiv& \sqrt{-m^2/\lambda}.
\eeq
This is used in many works, because it has the advantage that $v_{\rm tree}$ is manifestly 
independent of the choice of gauge-fixing. However, it has the disadvantage
that Higgs tadpole loop diagrams do not vanish, and must be included order-by-order in 
perturbation theory. This comes with a parametrically slower convergence of perturbation 
theory, as the tadpole contributions to other calculated quantities will
include powers of $1/\lambda$ due to their
zero-momentum Higgs propagators.

We choose instead to expand the Higgs field around a loop-corrected VEV $v$,
which is defined to be the minimum of the full effective potential 
\cite{Coleman:1973jx,Jackiw:1974cv,Sher:1988mj}
in Landau gauge. For the Standard Model (and indeed for a general renormalizable
field theory), the effective potential has now been obtained at 2-loop  
\cite{Ford:1992pn, Martin:2001vx} 
and 3-loop 
\cite{Martin:2013gka, Martin:2017lqn} 
orders, with the 4-loop contributions known \cite{Martin:2015eia}
at leading order in QCD. The choice of Landau gauge is made because
other gauge-fixing choices lead to unpleasant
technical problems including
kinetic mixing between the longitudinal components of the vector 
and the Goldstone scalar degrees of 
freedom.\footnote{The full 2-loop effective potential
has been recently obtained in a large class of more general gauge-fixing schemes in 
ref.~\cite{Martin:2018emo}, but it is quite unwieldy, and extending it 
to 3-loop order is a daunting challenge.}
The disadvantage of defining the VEV in this way is that 
calculations that make use of it are
then restricted to Landau gauge. But the advantage of this choice is that
the sum of all Higgs tadpole diagrams (including the tree-level tadpole) automatically
vanishes, and there are no corresponding $1/\lambda^n$ contributions in 
perturbation theory.
 
Another issue to be dealt with is that the minimization condition
for the effective potential requires resummation 
of Goldstone boson contributions, 
as explained in \cite{Martin:2014bca,Elias-Miro:2014pca},
in order to avoid spurious imaginary parts and infrared divergences 
at higher loop orders.
(For further perspectives and developments on this issue, see refs.~\cite{Pilaftsis:2015cka,
Pilaftsis:2015bbs,
Kumar:2016ltb,
Espinosa:2016uaw,
Braathen:2016cqe,
Pilaftsis:2017enx,
Braathen:2017izn}.)
The end result can be written as a relation between the tree-level and
loop-corrected VEVs:
\beq
v^2_{\rm tree} = v^2 + 
\frac{1}{\lambda} \sum_{n = 1}^\infty
\frac{1}{(16\pi^2)^n} \Delta_n
,
\label{eq:Veffmincon}
\eeq
with $n$-loop order contributions $\Delta_n$ that are free of spurious imaginary parts and infrared divergences and do not depend at all on the Goldstone boson squared mass. 
(The $1/\lambda$ in this
equation is the source of the tadpole effects noted above if one chooses to
expand in terms of $v_{\rm tree}$ rather than $v$.)
The full 3-loop contributions were given in \cite{Martin:2017lqn} in terms of 2-loop and 3-loop
basis integrals that can be efficiently evaluated numerically using the
computer code {\tt 3VIL} \cite{Martin:2016bgz},\footnote{{\tt 3VIL} computes 3-loop vacuum basis 
integrals numerically using the differential equations method, except in special cases for which 
they can be computed analytically, including the cases found in 
refs.~\cite{Chetyrkin:1981qh}-\cite{Burda:2017tcu}. See ref.~\cite{TVID} for an 
alternative evaluation of 3-loop vacuum integrals based on dispersion relations.}
and the 4-loop contribution was obtained
at leading order in QCD in \cite{Martin:2015eia}. However, a numerical 
illustration of these effects was deferred. One of the purposes of the present 
paper is to remedy this by providing a numerical study of the 3-loop and 4-loop
effects.

We also have a broader purpose here; to bring together in a coherent form,
implemented as a public computer code, results obtained in recent years 
relating pole masses and other observables to the Lagrangian parameters
in the tadpole-free pure \MSbar scheme. The new code, called {\tt SMDR} for
Standard Model in Dimensional Regularization, is a software
library written in C with functions 
callable from user C or C++ programs.
It uses the \MSbar input parameters
that define\footnote{Cabibbo-Kobayashi-Maskawa mixing 
and neutrino mass and mixing effects
are neglected in the present version. Including them would have a negligible
effect on the quantities in eq.~(\ref{eq:onshellinputs}), compared to other sources of uncertainty.} 
the Standard Model theory
at a given renormalization scale $Q$:
\beq
&&v,\> \lambda,\> g_3,\> g,\> g',\> y_t,\> y_b,\> y_c,\> y_s,\> y_d,\> y_u,\> y_\tau,\> y_\mu,\> y_e,\>\Delta \alpha^{(5)}_{\rm had}(M_Z).
\label{eq:MSbarinputs}
\eeq 
All of these, except the last, 
are defined as running parameters in the non-decoupled (high-energy)
Standard Model, with gauge group $SU(3)_c \times SU(2)_L \times U(1)_Y$
with gauge couplings $g_3$, $g$, and $g'$ respectively,
and 6 active quarks.
Note that the running \MSbar Higgs squared mass parameter
$m^2$ need not be included among these,
because it is not independent, being determined in terms of $\lambda$, $v$,
and the other parameters
by the effective potential
minimization condition eq.~(\ref{eq:Veffmincon}). 
Also, the hadronic light-quark 
contribution to the fine-structure constant is given by a parameter 
$\Delta \alpha^{(5)}_{\rm had}(M_Z)$. In principle this is not independent
of the others in eq.~(\ref{eq:MSbarinputs}), 
but in practice it must (at least, at present) be treated as an independent input 
because it depends on non-perturbative physics.
The code then provides computations of the
following ``on-shell" output quantities:
\beq
\mbox{heavy particle pole masses:}&& M_t,\> M_h,\> M_Z,\> M_W,
\nonumber
\\
\mbox{running light quark masses:}&&m_b(m_b),\> m_c(m_c),\> m_s(\mbox{2 GeV}),\>
m_d(\mbox{2 GeV}),\>
m_u(\mbox{2 GeV}),\phantom{xxx}
\nonumber
\\
\mbox{lepton pole masses:}&& M_\tau,\> M_\mu,\> M_e,
\nonumber
\\
\mbox{5-quark QCD coupling:}&&\alpha_S^{(5)}(M_Z),
\nonumber
\\
\mbox{Fermi constant:}&& G_F = 1.1663787\ldots \times 10^{-5}\> \mbox{GeV}^{-2},
\nonumber
\\
\mbox{fine structure constant:}&&\alpha_0 = 1/137.035999139\ldots
\>\mbox{and}\>\>\Delta \alpha^{(5)}_{\rm had}(M_Z)
,
\label{eq:onshellinputs}
\eeq
which can be viewed as dual to the \MSbar inputs. 
(Even though $G_F$ and $\alpha_0$ are extremely accurately 
known from experiment, as indicated,
they are considered as outputs from the point of view of the pure \MSbar
renormalization scheme.)
However, note that $M_W$ is actually extra, in the sense that the other 
parameters in eq.~(\ref{eq:onshellinputs})
are already sufficient to fix the \MSbar quantities in eq.~(\ref{eq:MSbarinputs});
therefore, the computation of $M_W$ provides a consistency 
check on the Standard Model.
The quantity $\Delta \alpha^{(5)}_{\rm had}(M_Z)$ appears in both lists
(\ref{eq:MSbarinputs}) and (\ref{eq:onshellinputs}),
due to its non-perturbative nature; it always is obtained from experiment 
rather than fits to other quantities. 
The {\tt SMDR} code also computes the 
weak mixing angle as defined by the Particle Data Group's Review of Particle Properties (RPP) 
\cite{RPP} (which, unlike the present paper, uses a scheme with the top quark decoupled but the 
massive $W$ boson active, corresponding to a non-renormalizable effective theory 
even when the Lagrangian couplings of negative mass dimension are neglected), 
but this is again extra, since it is not needed in order to fix the \MSbar quantities. 

The relationship between the Sommerfeld fine-structure constant $\alpha_0$ appearing in eq.~(\ref{eq:onshellinputs})
and the couplings $g$ and $g'$ in eq.~(\ref{eq:MSbarinputs}) can be expressed as
(see, for example, refs.~\cite{Fanchiotti:1992tu,Erler:1998sy,Degrassi:2003rw,Degrassi:2014sxa}):
\beq
\alpha_0 =  \frac{g^2(M_Z) g^{\prime 2}(M_Z)}{4 \pi \left [g^2(M_Z) + g^{\prime 2}(M_Z) \right ]} \left [
1 
-\Delta \alpha^{(5)}_{\rm had}(M_Z) 
-\Delta \alpha_{\rm pert}^{\rm LO} 
-\Delta \alpha_{\rm pert}^{\rm HO}
\right ],
\eeq
where the sum of 1-loop contributions from $t,W,\tau,\mu,e$ (but not $b,c,s,d,u$) are: 
\beq
\Delta \alpha_{\rm pert}^{\rm LO} &=& \frac{\alpha_0}{4 \pi} 
\biggl [
\frac{202}{27} + 14 \ln(M_W/M_Z) - \frac{32}{9} \ln(M_t/M_Z)
- \frac{8}{3} \ln(M_\tau/M_Z) 
\nonumber \\ &&
- \frac{8}{3} \ln(M_\mu/M_Z) - \frac{8}{3} \ln(M_e/M_Z)
\biggr ] 
,
\eeq
and the higher-order perturbative contribution
$\Delta \alpha_{\rm pert}^{\rm HO}$ has been given as an interpolating formula
in eqs.~(19)-(21) of ref.~\cite{Degrassi:2014sxa}. For the running $\alpha^{\overline{\rm MS}}(Q)$ in the
decoupled theories used for renormalization group 
(RG) running below $M_Z$ [with the numbers of active
(quarks, charged leptons) equal to (5,~3) or (4,~3) or (4,~2) or (3,~2)], 
we use the results obtained in \cite{Martin:2018yow}, as discussed in the next section.

The pole masses $M_t$, $M_h$, $M_Z$, $M_W$, $M_\tau$, $M_\mu$, and $M_e$
are each defined in terms of the complex pole in the renormalized propagator, 
\beq
s_{\rm pole} = M^2 - i \Gamma M.
\label{eq:spole}
\eeq
For the top-quark pole mass, the pure QCD contributions were obtained at  
1-loop, 2-loop, 3-loop, and 4-loop orders in 
refs.~\cite{Tarrach:1980up},
\cite{Gray:1990yh}, 
\cite{Melnikov:2000qh}, and 
\cite{Marquard:2015qpa,Marquard:2016dcn}, 
respectively. The non-QCD contributions to $M_t$ at 1-loop and 2-loop orders had also 
been obtained in other schemes and approximations. 
At 1-loop order they were found in 
refs.~\cite{Bohm:1986rj,Hempfling:1994ar,Jegerlehner:2002em}, and mixed electroweak-QCD 2-loop
contributions were obtained in \cite{Jegerlehner:2003py,Eiras:2005yt,Jegerlehner:2012kn}. 
Further 2-loop contributions in the gauge-less limit (in which the electroweak boson masses
are taken to be small compared to the top-quark mass) were found in
refs.~\cite{Faisst:2003px,Jegerlehner:2003sp,Faisst:2004gn,Kniehl:2014yia}. 
Finally, the full 2-loop results for $M_t$ 
were provided in the tree-level VEV scheme 
in ref.~\cite{Kniehl:2015nwa}, and in the tadpole-free scheme used in the present paper in
\cite{Martin:2016xsp}.
 
For the Higgs boson mass, we use our calculation in ref.~\cite{Martin:2014cxa}, which contains all
2-loop contributions and the leading (in the limit $g^2,g^{\prime 2},\lambda \ll g_3^2, y_t^2$)
3-loop contributions in the tadpole-free pure \MSbar scheme.
Earlier works on $M_h$ at the 2-loop level
in other schemes and approximations include ref.~\cite{Bezrukov:2012sa}
which included the mixed QCD/electroweak contributions to $M_h$,
ref.~\cite{Degrassi:2012ry} which used the gauge-less limit approximation
at 2-loop order, and the full 2-loop approximation given as an interpolating
formula in a hybrid \MSbar/on-shell scheme in ref.~\cite{Buttazzo:2013uya}.

For the $W$ and $Z$ boson pole masses, we use the full 2-loop calculations 
using the tadpole-free pure \MSbar scheme
given in refs.~\cite{Martin:2015lxa} and \cite{Martin:2015rea}, respectively.
Previous 2-loop calculations of the vector boson pole masses in other schemes
(expanding around $v_{\rm tree}$ rather than $v$) 
appeared in refs.~\cite{Jegerlehner:2001fb}, \cite{Jegerlehner:2002em}, 
\cite{Degrassi:2014sxa}, and \cite{Kniehl:2015nwa}. 
It is important to note that for the vector bosons $V=W$ and $Z$, the values usually quoted,
including by the RPP, are
not the pole masses but the variable-width Breit-Wigner masses. These can be related to the
pole masses by \cite{Bardin:1988xt,Willenbrock:1991hu,Sirlin:1991fd,Stuart:1991xk}: 
\beq
M^2_{V,\,\small \mbox{Breit-Wigner}} = M_V^2 + \Gamma_{V}^2.
\eeq
Thus, the $Z$- and $W$-boson pole masses defined by eq.~(\ref{eq:spole}) are, respectively, 
approximately 34.1 MeV and 27.1 MeV
smaller than the Breit-Wigner masses that are usually quoted.

The charged lepton pole masses are computed at 2-loop order in QED, by converting the corresponding 
QCD formulas given in ref.~\cite{Gray:1990yh} and including 
small effects from non-zero lighter fermion masses from ref.~\cite{Bekavac:2007tk}.

The running light-quark masses in eq.~(\ref{eq:onshellinputs})
are defined in appropriate $SU(3)_c \times U(1)_{\rm EM}$ effective
field theories in which the heavier particles have been decoupled.
Although it is possible to evaluate the QCD contributions 
to the bottom-quark and charm-quark pole masses, this is deprecated, 
because there is no semblance of convergence of the perturbative series relating
the pole masses to the running masses for bottom and charm (and obviously for the lighter
quarks as well); see ref.~\cite{Marquard:2016dcn}.
Therefore we use running \MSbar masses for all lighter quarks.
Thus $m_b(m_b)$ is defined as an \MSbar running mass
in the 5-quark, 3-lepton QCD+QED effective theory,
while $m_c(m_c)$ is similarly defined in the 4-quark, 2-lepton theory,
and $m_s(\mbox{2 GeV}),\>
m_d(\mbox{2 GeV}),\>
m_u(\mbox{2 GeV})$ are defined in the 3-quark, 2-lepton theory. We follow
the RPP ref.~\cite{RPP} in choosing to evaluate the last three at, somewhat arbitrarily,
$Q = 2$ GeV, in order to avoid larger QCD effects at smaller $Q$.

To obtain the 5-quark, 3-lepton QCD+QED effective field theory, we simultaneously
decouple the heavier Standard Model particles $t,h,Z,W$ at a common
matching scale, which can be chosen at will, but should presumably be in the
range from about $M_W$ to $M_t$. Because $W$ and $Z$ are decoupled from it, 
this low-energy effective theory is a renormalizable gauge theory supplemented by interactions
with couplings of
negative mass dimension (including the Fermi four-fermion interactions).
The decouplings of the bottom quark, tau lepton, and charm quark are then performed individually.

In one mode of operation, the {\tt SMDR} code takes 
the \MSbar input parameters of eq.~(\ref{eq:MSbarinputs}) provided by the user,
and outputs the on-shell quantities in eq.~(\ref{eq:onshellinputs}).
Alternatively, in a dual mode of operation, 
the {\tt SMDR} code instead takes user input for the on-shell quantities
in eq.~(\ref{eq:onshellinputs}) (except for $M_W$), and determines as outputs the
\MSbar quantities in eq.~(\ref{eq:MSbarinputs}) and then $M_W$, by doing a fit.
The {\tt SMDR} code also implements all known contributions to the running and
decoupling of the gauge and Yukawa couplings.

In the numerical studies below, we employ a benchmark 
model point, chosen to yield the central values of the quantities in eq.~(\ref{eq:onshellinputs}) (other than $M_W$, as noted above), as given in the 2019 update of the 2018 
edition of the Review of Particle Properties ref.~\cite{RPP}:
\beq
&&
M_t \>=\> \mbox{173.1 GeV},\qquad 
M_h \>=\> \mbox{125.1 GeV},\qquad
M_{Z,\,\small \mbox{Breit-Wigner}} \>=\> \mbox{91.1876 GeV},
\nonumber 
\\
&&
G_F \>=\> 1.1663787 \times 10^{-5} \> {\rm GeV}^2,
\qquad
\alpha_0 \>=\> 1/137.035999139,
\qquad
\alpha_S^{(5)}(M_Z) = 0.1181,
\nonumber
\\
&&
m_b(m_b) \>=\> \mbox{4.18 GeV},
\qquad
m_c(m_c) \>=\> \mbox{1.27 GeV},
\qquad
m_s(\mbox{2 GeV}) \>=\> \mbox{0.093 GeV} 
\nonumber
\\
&&
m_d(\mbox{2 GeV}) \>=\> \mbox{0.00467 GeV},
\qquad 
m_u(\mbox{2 GeV}) \>=\> \mbox{0.00216 GeV},
\qquad
M_\tau \>=\> \mbox{1.77686 GeV},
\nonumber
\\
&&
M_\mu \>=\> \mbox{0.1056583745 GeV},
\qquad
M_e \>=\> \mbox{0.000510998946 GeV},
\nonumber
\\
&& 
\Delta \alpha^{(5)}_{\rm had}(M_Z) \>=\> 0.02764,
\label{eq:referencemodelOS}
\eeq
The \MSbar input quantities that do this are found (with default scale choices for evaluations in {\tt SMDR}) to be:
\beq
Q_0 &=& 173.1\>{\rm GeV},
\nonumber
\\
v(Q_0) &=& 246.60109\>{\rm GeV},
\qquad 
\lambda(Q_0) \>=\> 0.12603842,
\nonumber
\\
g_3(Q_0) &=& 1.1636241,\qquad\>\>\, 
g_2(Q_0) \>=\> 0.64765961,\qquad\> 
g'(Q_0) \>=\>  0.35853877,
\nonumber
\\
y_t(Q_0) &=&   0.93480082,\qquad
y_b(Q_0) \>=\> 0.015480097,\qquad
y_\tau(Q_0) \>=\> 0.0099944422,
\nonumber
\\
y_c(Q_0) &=& 0.0033820038, 
\qquad
y_s(Q_0) \>=\> 0.00029094484,
\qquad
y_\mu(Q_0)  \>=\> 0.00058837986,\phantom{xxx}
\nonumber
\\
y_d(Q_0) &=&   1.4609792 \times 10^{-5},
\qquad
y_u(Q_0) \>=\> 6.7227779\times 10^{-6},
\nonumber
\\
y_e(Q_0) &=&   2.7929820 \times 10^{-6}.
\label{eq:referencemodelMSbar}
\eeq
This set of values obviously includes more significant digits
than justified by the experimental and theoretical uncertainties; this is for the sake
of reproducibility and checking when changes are made to the code, or to the 
default choices of matching or evaluation scales. Equation (\ref{eq:referencemodelMSbar}) 
will be referred to below as the reference model point,
and a sample input file included with the {\tt SMDR} distribution provides for 
automatic loading of these parameters. As future versions of the RPP
with new experimental results become available, corresponding 
new versions of the reference model file will be included in new {\tt SMDR}
distributions; they can also be constructed easily by using functions provided.
All of the figures appearing below are made using short programs 
(included with the {\tt SMDR} distribution) that employ 
the {\tt SMDR} library functions,
in order to illustrate how the latter should be used.

\section{Renormalization group running and decoupling\label{sec:RGEs}}
\setcounter{equation}{0}
\setcounter{figure}{0}
\setcounter{table}{0}
\setcounter{footnote}{1}

The \MSbar renormalization group equations for the Standard Model used in this paper, 
and by default in the
{\tt SMDR} code, are the state-of-the-art ones. These include the 2-loop 
\cite{MVI,MVII,Jack:1984vj,MVIII,Luo:2002ey} and 3-loop
\cite{Tarasov,Mihaila:2012fm,Chetyrkin:2012rz,Bednyakov:2012rb,Bednyakov:2012en,
Chetyrkin:2013wya,Bednyakov:2013eba,Bednyakov:2013cpa,Bednyakov:2014pia}
order contributions for all parameters, including the gauge couplings, the fermion Yukawa couplings,
the Higgs self-coupling $\lambda$, VEV $v$, and negative squared mass $m^2$. In addition, 
for the strong coupling, the contributions to the beta function 
at 4-loop order in the limit $g^2, g^{\prime 2} \ll g_3^2, y_t^2, \lambda$
\cite{vanRitbergen:1997va,Czakon:2004bu,
Bednyakov:2015ooa,Zoller:2015tha,Poole:2019txl} 
and pure QCD 5-loop order 
\cite{Baikov:2016tgj,Herzog:2017ohr} are included. 
Similarly, the higher-order QCD contributions to the beta functions of the quark Yukawa couplings
are included, using results found at 4-loop order in 
refs.~\cite{Chetyrkin:1997dh,Vermaseren:1997fq} and at 5-loop order in ref.~\cite{Baikov:2014qja}.
Finally, the leading QCD 4-loop contribution to the beta function of 
the Higgs self-coupling $\lambda$ is included from refs.~\cite{Martin:2015eia,Chetyrkin:2016ruf}.

\begin{figure}[!p]
  \begin{minipage}[]{0.495\linewidth}
    \includegraphics[width=8.0cm,angle=0]{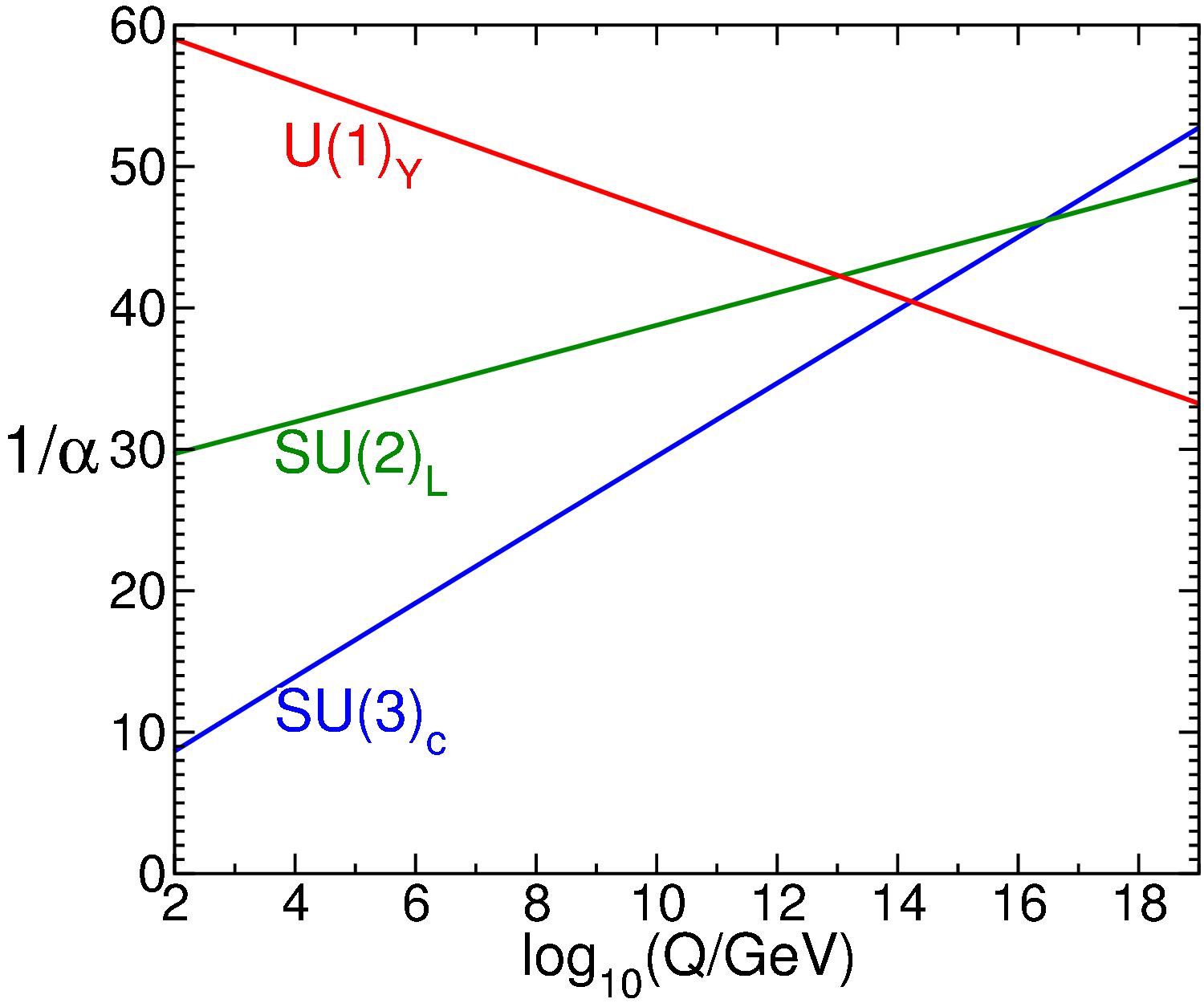}
  \end{minipage}
 \begin{minipage}[]{0.495\linewidth}
    \includegraphics[width=8.0cm,angle=0]{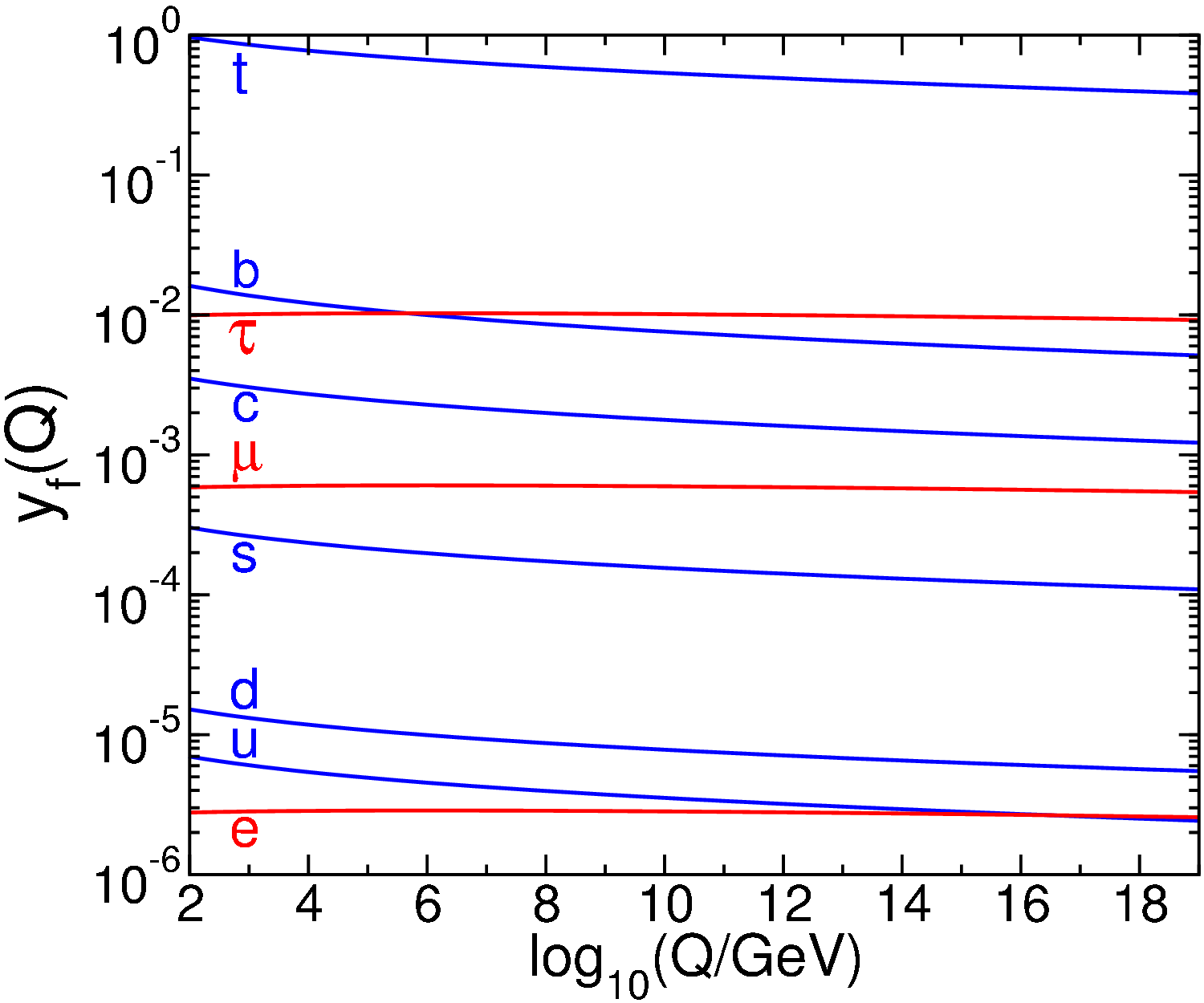}
  \end{minipage}
\begin{center}\begin{minipage}[]{0.95\linewidth}
\caption{\label{fig:aboveMZ}Renormalization group running of the \MSbar inverse
gauge couplings $1/\alpha_3$, $1/\alpha_2$, and $1/\alpha_1$ in a grand unified theory normalization
(left panel) and charged
fermion Yukawa couplings (right panel), as functions of the renormalization scale $Q$.
The input parameters are given by the reference model point
defined in eq.~(\ref{eq:referencemodelMSbar}) at $Q_0 = 173.1$ GeV.}
\end{minipage}\end{center}
\vspace{1.5cm}
  \begin{minipage}[]{0.495\linewidth}
    \includegraphics[width=8.0cm,angle=0]{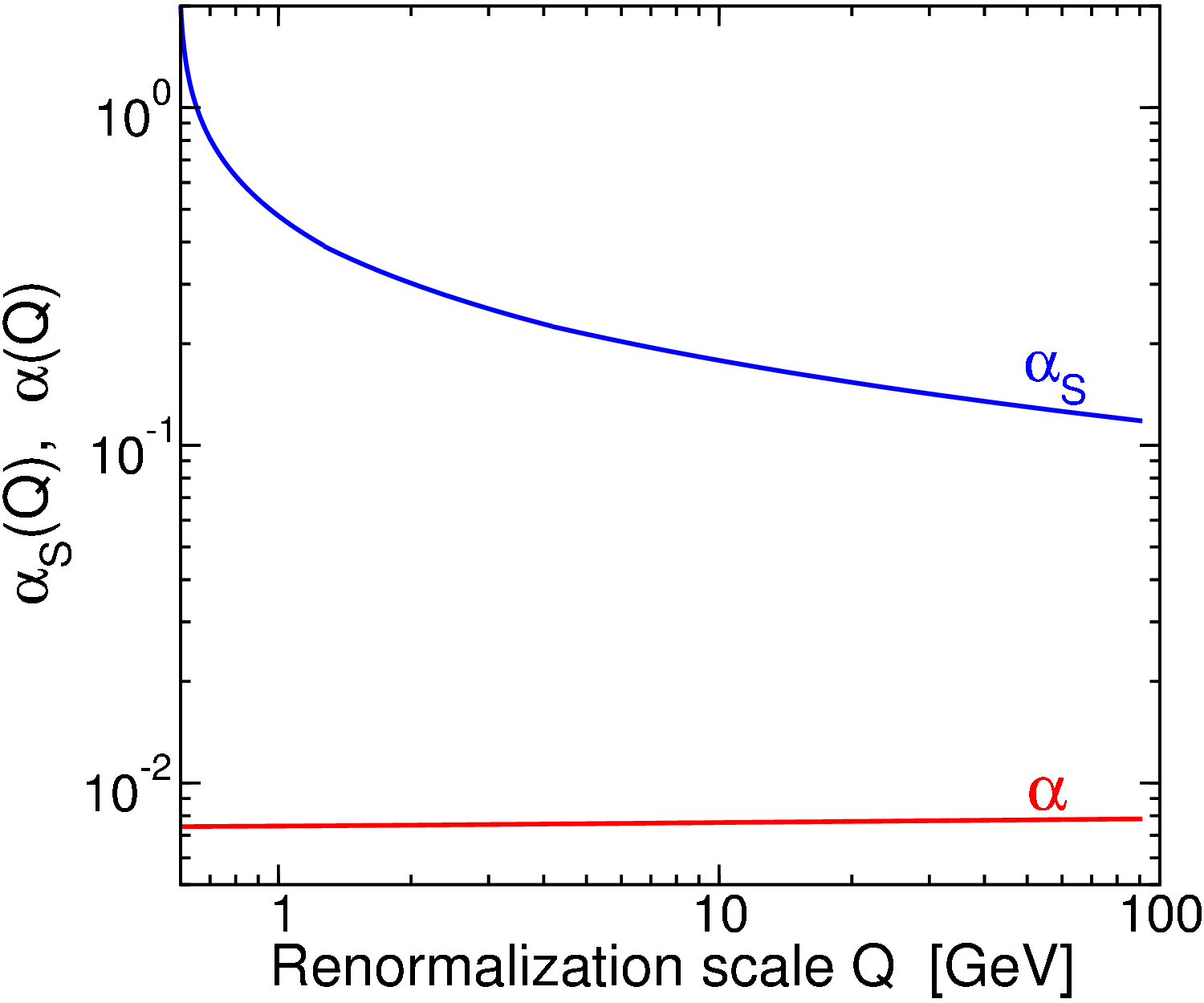}
  \end{minipage}
 \begin{minipage}[]{0.495\linewidth}
    \includegraphics[width=8.0cm,angle=0]{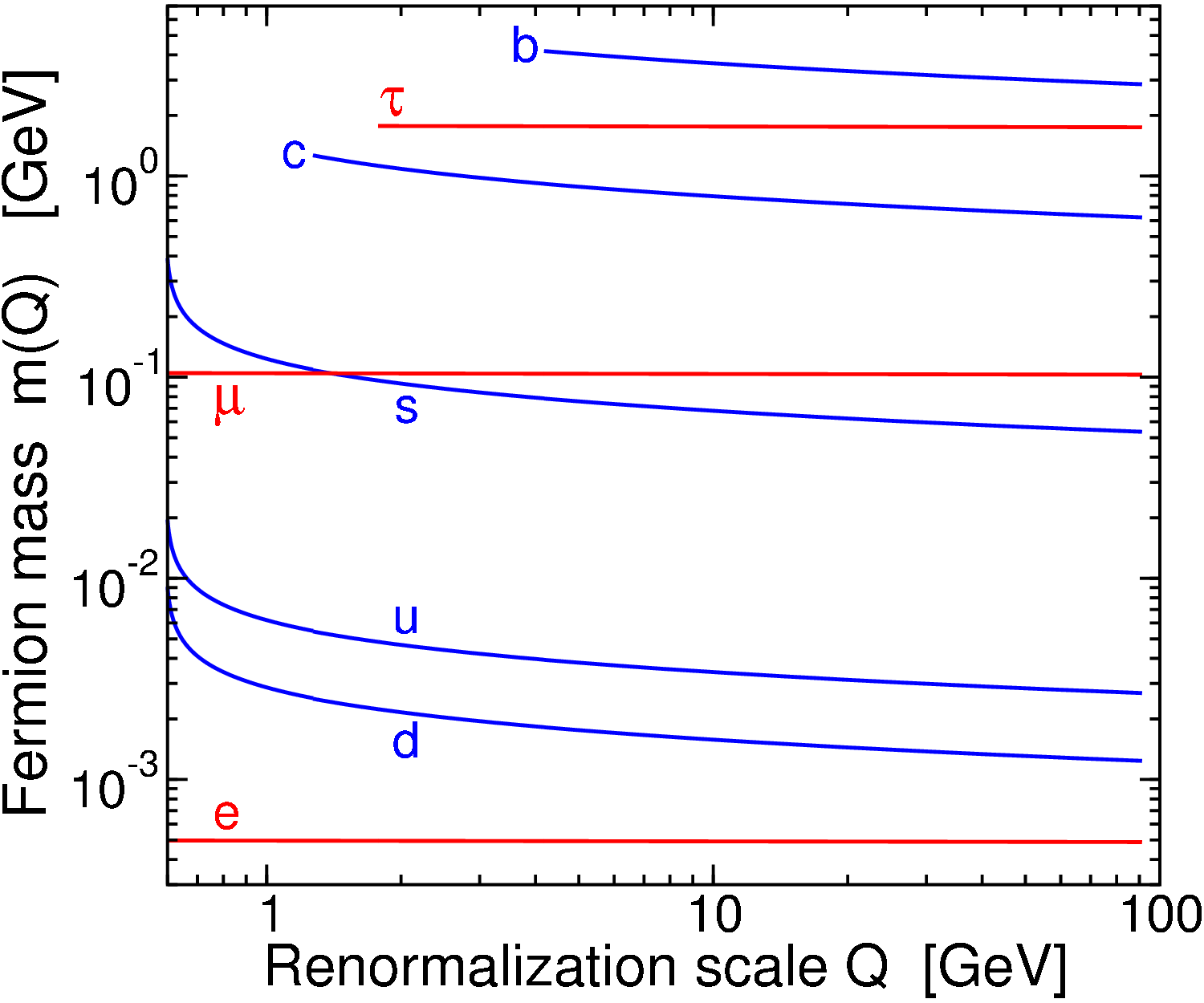}
  \end{minipage}
\begin{center}\begin{minipage}[]{0.95\linewidth}
\caption{\label{fig:belowMZ}Renormalization group running of the \MSbar QCD and QED gauge couplings 
$\alpha_S$ and $\alpha$ (left panel) and fermion masses (right panel), 
as functions of the renormalization scale $Q$.
The beta functions used are 5-loop order in QCD and 3-loop order in QED, 
with active fermion contents as follows:
5-quark, 3-lepton  for $m_b(m_b) \leq Q \leq \mbox{91.1876 GeV}$;
4-quark, 3-lepton for $M_\tau \leq Q \leq m_b(m_b)$;
4-quark, 2-lepton for $m_c(m_c) \leq Q \leq M_\tau$;
and 3-quark, 2-lepton for $Q \leq m_c(m_c)$. 
The matchings at $Q = m_b(m_b)$ and $M_\tau$ and $m_c(m_c)$ are 
done at 4-loop order for the QCD coupling, 2-loop order for the QED coupling,
and the fermion mass matchings include effects at 3-loop order in QCD and 2-loop order in QED.
The input parameters are defined by the reference model point given in 
eq.~(\ref{eq:referencemodelMSbar}),
with $t,h,Z,W$ simultaneously decoupled at $Q =91.1876$ GeV.}
\end{minipage}\end{center}
\end{figure}

Using the reference model of eq.~(\ref{eq:referencemodelMSbar}) as inputs, the renormalization group
running of the couplings are illustrated in Figure \ref{fig:aboveMZ} for the range
$10^2$ GeV $< Q < 10^{19}$ GeV. The left panel
shows the inverse gauge couplings $1/\alpha_3 = 4 \pi/g_3^2$ and $1/\alpha_2 = 4 \pi/g^2$
and (in a Grand Unified Theory [GUT] normalization) $1/\alpha_1 = (3/5) 4 \pi/g^{\prime 2}$,
while the right panel shows the Yukawa couplings for all of the Standard Model charged fermions.

For lower scales, we use the results given in ref.~\cite{Martin:2018yow} to simultaneously decouple
the top quark, Higgs boson, $Z$ boson, and $W$ boson at a common matching scale, so that the 
low-energy effective field theory is renormalizable and has gauge group $SU(3)_c \times U(1)_{\rm EM}$. 
The common matching scale is, in principle, arbitrary; by 
default the {\tt SMDR} code uses $Q=M_Z$ for the
matching but this can be modified at run time by the 
user. The matching results include
the 2-loop matching found in \cite{Martin:2018yow} 
for the electromagnetic \MSbar coupling $\alpha(Q)$ 
in the theory with 5 quarks and 3 leptons, as well as the matching relation 
for the 5-quark QCD coupling
$\alpha_S(Q)$ at 1-loop \cite{Weinberg:1980wa,Ovrut:1980dg}, 
2-loop \cite{Bernreuther:1981sg,Larin:1994va}, 
3-loop \cite{Chetyrkin:1997un,Grozin:2011nk}, 
and 4-loop \cite{Schroder:2005hy,Chetyrkin:2005ia} orders together 
with the complete Yukawa and electroweak 2-loop contributions obtained first 
in ref.~\cite{Bednyakov:2014fua} 
(and verified and written in a different way compatible with the present paper
in ref.~\cite{Martin:2018yow}).
The pure QCD corrections to the quark mass matching relations were given at 3-loop order in ref.~\cite{Chetyrkin:1997un,Grozin:2011nk} and 4-loop order in ref.~\cite{Liu:2015fxa}.
 
For the QCD parts of the matching relations
and beta functions, complete results had 
been calculated and incorporated long ago into the {\tt RunDec} and {\tt CRunDec} 
\cite{Chetyrkin:2000yt,Schmidt:2012az,Herren:2017osy}
codes. 
In addition, the 2-loop mixed QCD/electroweak and pure electroweak contributions
to matching of the running $b,c,s,d,u$ and $\tau,\mu,e$ fermion masses were obtained in 
refs.~\cite{Kniehl:2004hfa,Kniehl:2014yia,Kniehl:2015nwa,Bednyakov:2016onn,Kniehl:2016enc}
and \cite{Martin:2018yow}. 
They are implemented in {\tt SMDR} using the formulas provided in
ref.~\cite{Martin:2018yow} consistent with the conventions of the present paper.

The running and decoupling of the QCD and QED gauge couplings and running 
fermion masses are shown in Figure \ref{fig:belowMZ} 
for the sequence of effective theories with 5 quarks and 3 charged leptons 
(for $m_b(m_b) \leq Q \leq M_Z$),
with 4 quarks and 3 charged leptons (for $M_\tau \leq Q \leq m_b(m_b)$),
with 4 quarks and 2 charged leptons (for $m_c(m_c) \leq Q \leq M_\tau$),
and with 3 quarks and 2 charged leptons (for $Q \leq m_c(m_c)$). The boundaries between these
effective theories are somewhat arbitrary, and correspond to the default
points within the {\tt SMDR} code, which can be adjusted by the user. At each of the matching points
$Q = m_b(m_b)$ and $M_\tau$ and $m_c(m_c)$, the parameters are actually discontinuous due to the
matching mentioned above due to changing effective theories, 
but this cannot be discerned with the resolution of the plots.

\clearpage

\section{Minimization of the effective potential and the
vacuum expectation value\label{sec:effpot}}
\setcounter{equation}{0}
\setcounter{figure}{0}
\setcounter{table}{0}
\setcounter{footnote}{1}

We first consider a numerical illustration of the minimization condition for the effective potential,
eq.~(\ref{eq:Veffmincon}),
which can be used to trade $m^2$ for $v$, 
when all of the other \MSbar parameters are taken to be known inputs. The quantities
$\Delta_n$ have been given up to 3-loop order in ref.~\cite{Martin:2017lqn} and
the 4-loop order contribution at leading order in QCD is found in 
ref.~\cite{Martin:2015eia}.

\begin{figure}[!p]
  \begin{minipage}[]{0.495\linewidth}
    \includegraphics[width=8.0cm,angle=0]{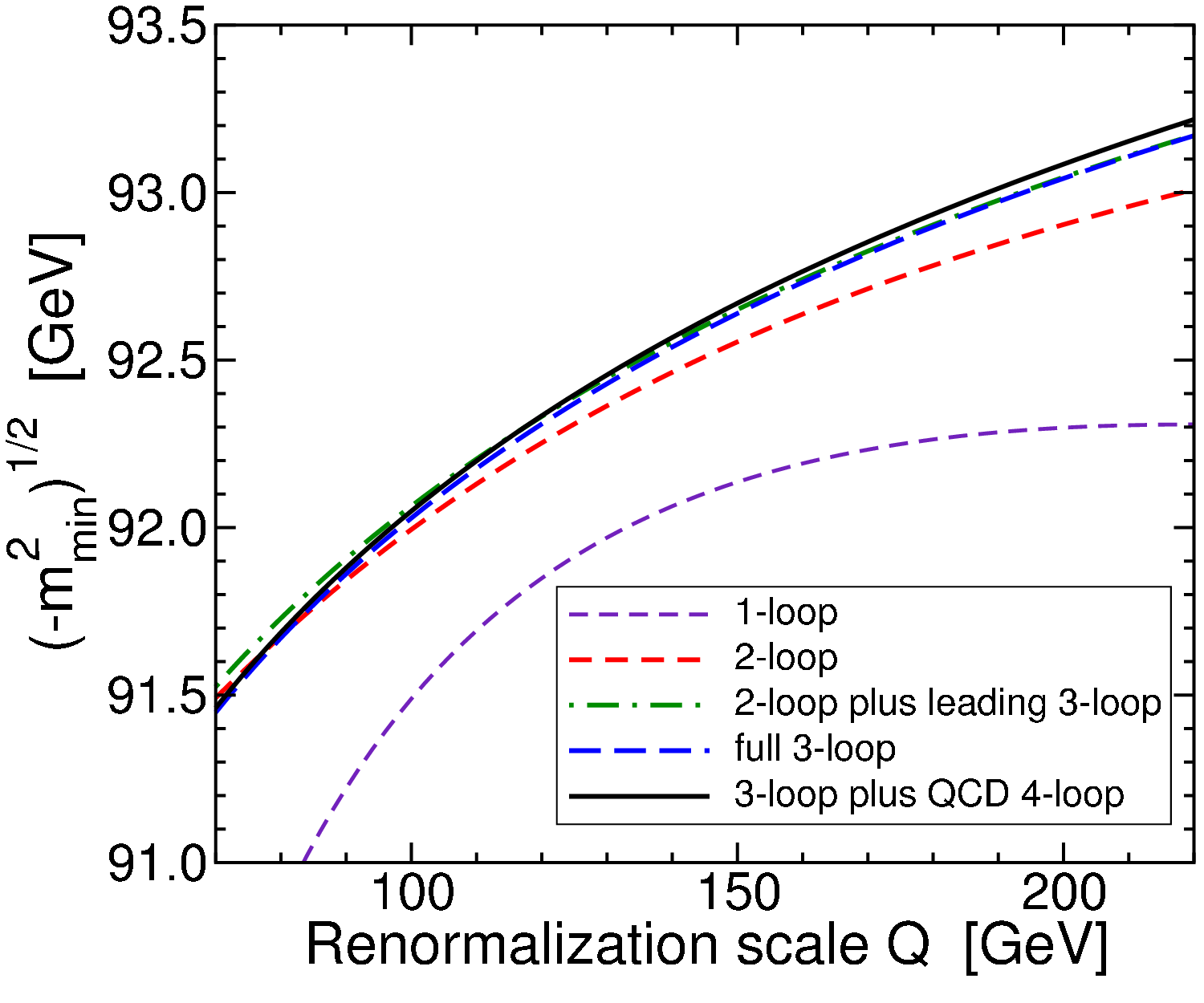}
  \end{minipage}
 \begin{minipage}[]{0.495\linewidth}
    \includegraphics[width=8.0cm,angle=0]{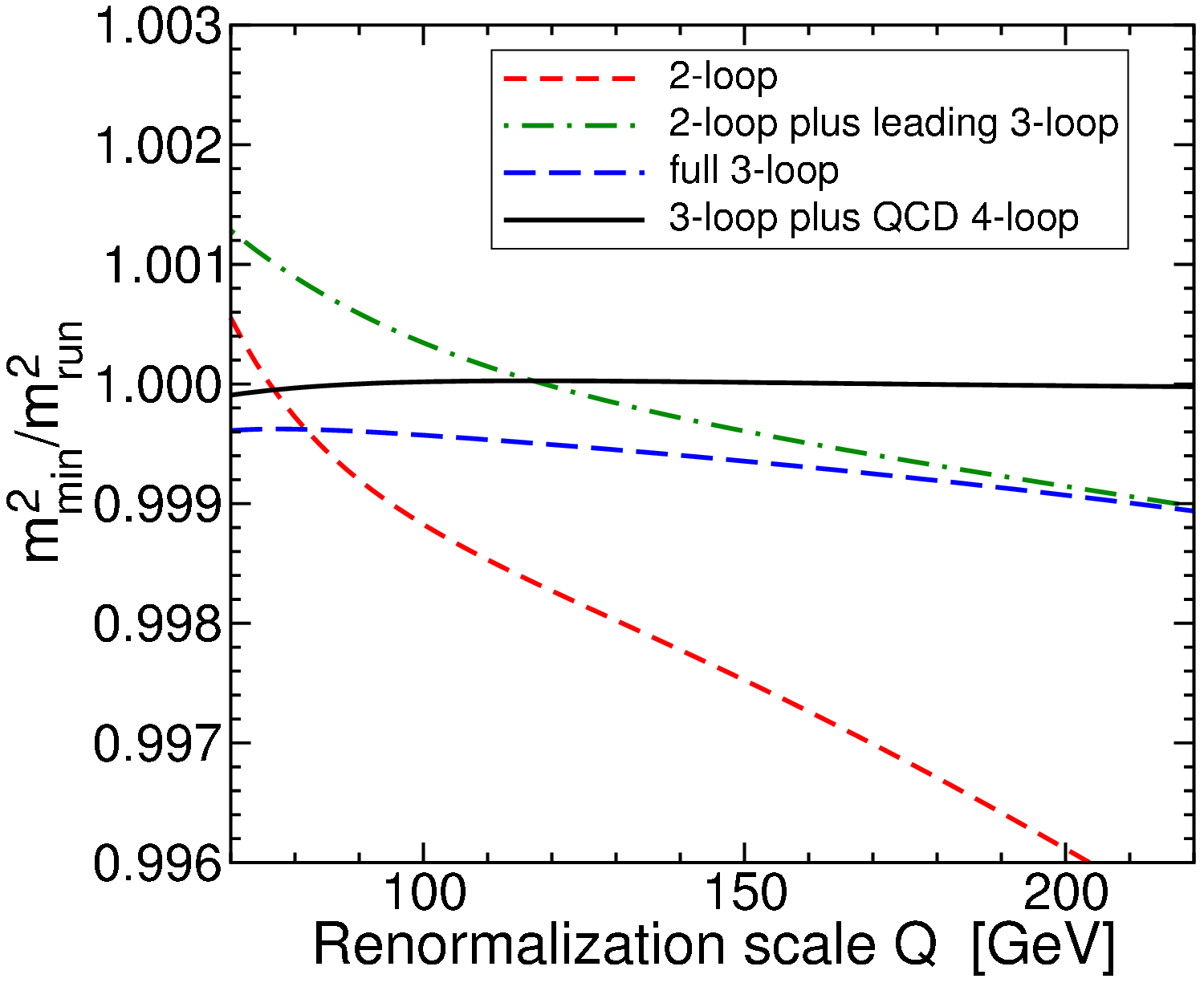}
  \end{minipage}
\begin{center}\begin{minipage}[]{0.95\linewidth}
\caption{\label{fig:m2}The \MSbar Higgs squared mass parameter,
as a function of the renormalization scale $Q$, for the reference model point
defined at $Q_0 = 173.1$ GeV in eq.~(\ref{eq:referencemodelMSbar}). 
The other input parameters, including the VEV $v(Q)$,
are obtained from the reference model by evolving them
using their RG equations to the scale $Q$, 
where the Landau gauge effective potential
is then required to be minimized to determine $m^2(Q)_{\rm{min}}$.  
In the left panel, results are shown for the 1-loop, 2-loop, 
2-loop plus leading 3-loop, full 3-loop, and 3-loop plus QCD 4-loop 
approximations to the effective potential minimization condition. 
The right panel shows the results for 
$m^2(Q)_{\rm{min}}/m^2(Q)_{\rm{run}}$, 
where $m^2(Q)_{\rm{min}}$ is determined as in the left panel, and $m^2(Q)_{\rm{run}}$ is obtained directly by 
renormalization running its input value 
from the reference scale $Q_0 = 173.1$ GeV.}
\end{minipage}\end{center}
\vspace{0.7cm}
  \begin{minipage}[]{0.495\linewidth}
    \includegraphics[width=8.0cm,angle=0]{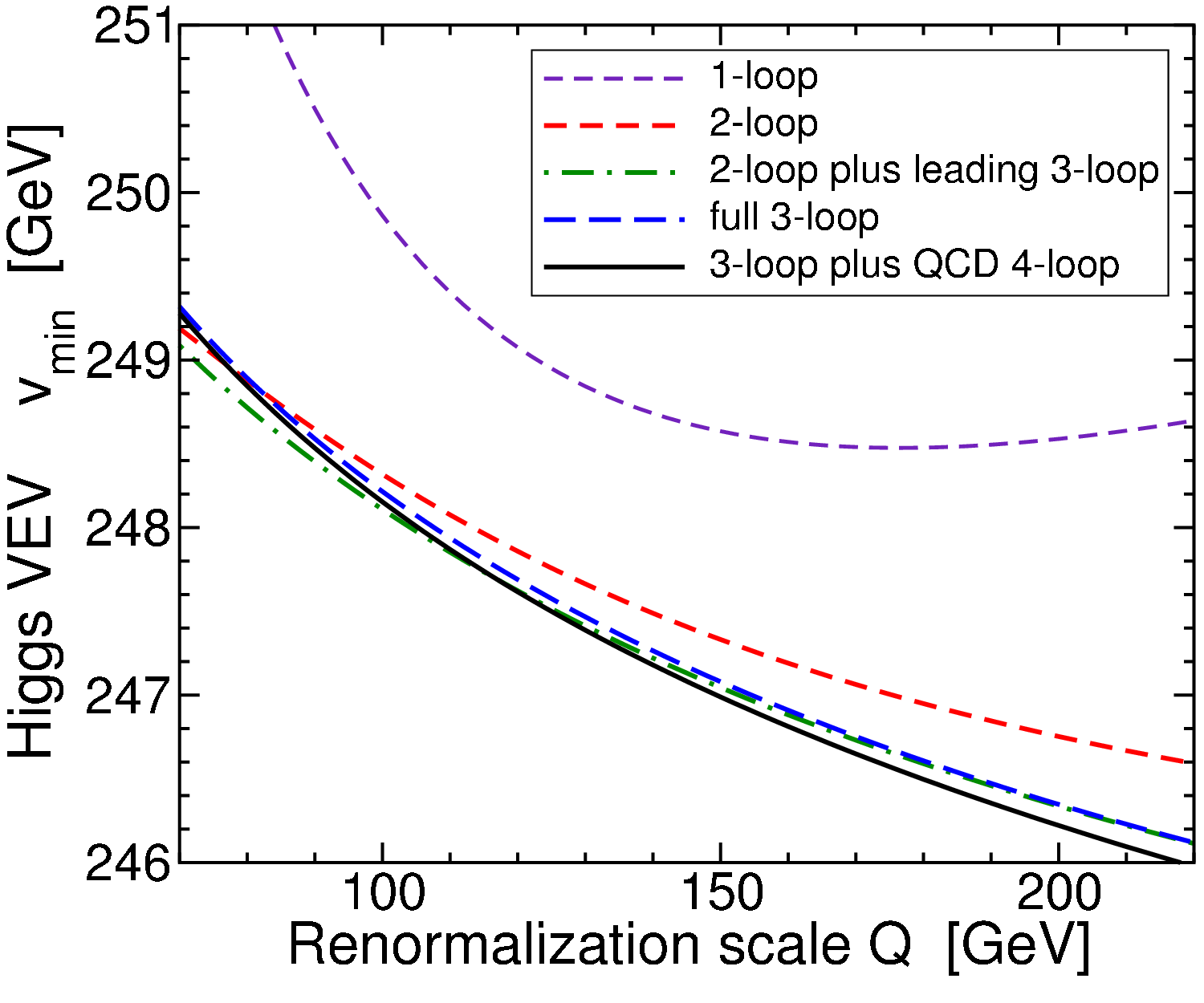}
  \end{minipage}
 \begin{minipage}[]{0.495\linewidth}
    \includegraphics[width=8.0cm,angle=0]{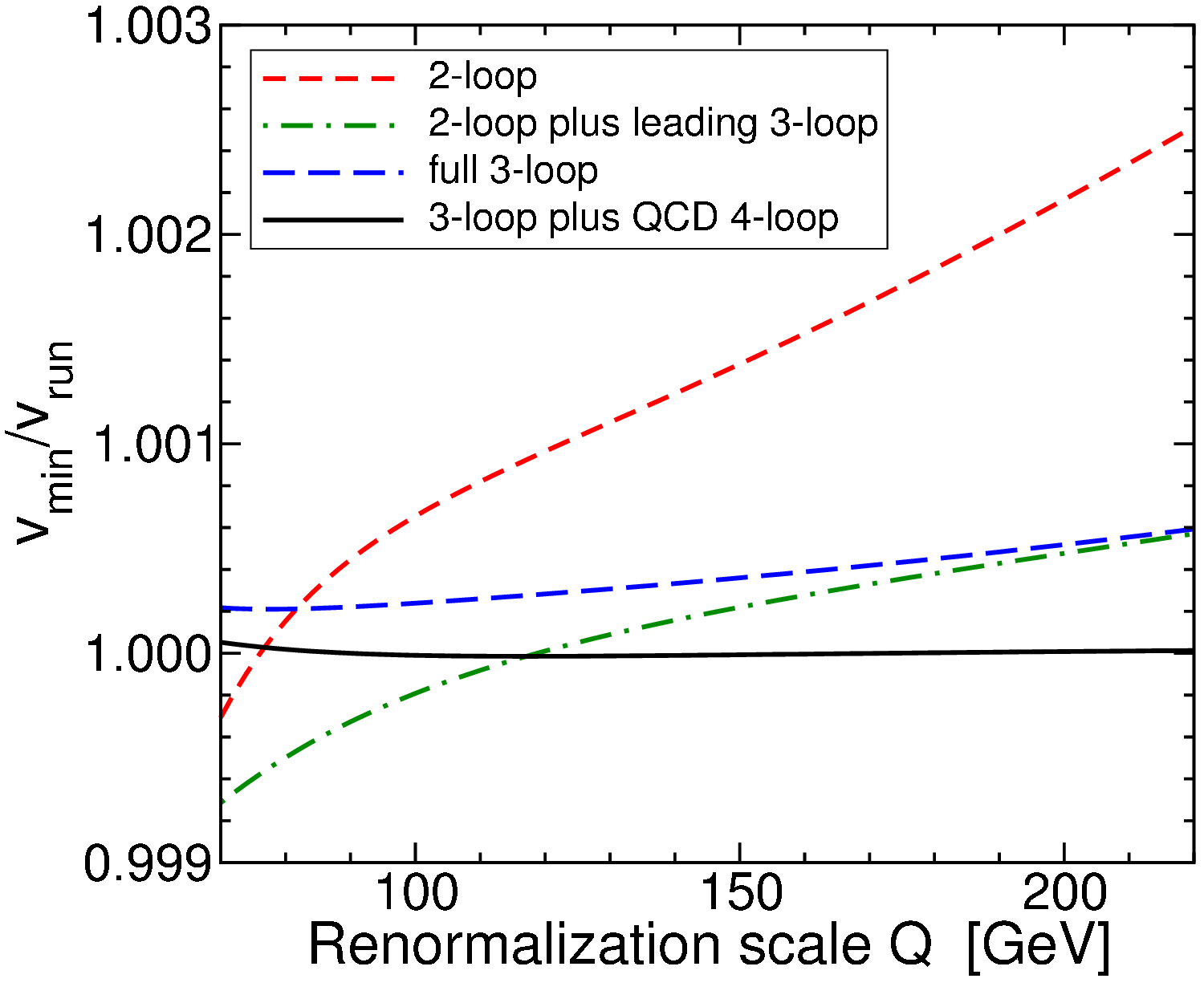}
  \end{minipage}
\begin{center}\begin{minipage}[]{0.95\linewidth}
\caption{\label{fig:vev} 
The \MSbar Higgs VEV, as a function of the renormalization scale $Q$, 
for the reference model point
defined at $Q_0 = 173.1$ GeV in eq.~(\ref{eq:referencemodelMSbar}). 
The other input parameters, including $m^2(Q)$,
are obtained from the reference model by evolving them
using their RG equations to the scale $Q$, 
where the Landau gauge effective potential
is minimized to obtain $v(Q)_{\rm{min}}$.  
In the left panel, results are shown for the 1-loop, 2-loop, 
2-loop plus leading 3-loop, full 3-loop, and 3-loop plus QCD 4-loop 
approximations to the effective potential minimization condition. 
The right panel shows the results for 
$v(Q)_{\rm{min}}/v(Q)_{\rm{run}}$, 
where $v(Q)_{\rm{run}}$ is obtained directly by 
renormalization running from the reference scale $Q_0 = 173.1$ GeV.}
\end{minipage}\end{center}
\end{figure}

In Figure \ref{fig:m2}, we start with the \MSbar quantities
taken to be their benchmark reference point values
defined at $Q = Q_0 = 173.1$ in eq.~(\ref{eq:referencemodelMSbar}).
From eq.~(\ref{eq:Veffmincon}), the value of $m^2$ at $Q_0$ 
for the reference model is then found to be
(again including more significant digits than justified by the uncertainties):
\beq
m^2(Q_0) &=& -(\mbox{92.878850 GeV})^2.
\label{eq:m2Qreference}
\eeq
At other renormalization group scales $Q$, we determine $m^2(Q)$ in two different ways. For the first way, we
renormalization-group 
run all of the other parameters to $Q$, where $m^2(Q)_{\rm min}$ 
is then determined by again applying eq.~(\ref{eq:Veffmincon}). 
The results are shown in the left panel of
Figure \ref{fig:m2}, in various approximations (as labeled) for the minimization condition.
The second way is to directly RG run $m^2(Q)_{\rm run}$ 
starting with eq.~(\ref{eq:m2Qreference}) as its boundary condition.
In the right panel, we show the ratio  of $m^2(Q)_{\rm min}/m^2(Q)_{\rm run}$ as
a function of $Q$.
This provides a scale-invariance check yielding a lower bound on the error, because 
in the idealized case of calculations to all orders 
in perturbation theory, the ratio should be exactly 1. 
We find that in the case of the full 3-loop plus QCD 4-loop approximation,
the deviation of the ratio from unity 
is less than $10^{-4}$ for the entire range shown
from 70 GeV to 220 GeV, 
and over most of this range the deviation is actually much smaller.
Without including the 4-loop QCD contribution, the scale dependence 
is still quite good, but is a few times $10^{-4}$. In both cases,
the parametric uncertainties from experimentally measured quantities
would seem to be probably larger than the theoretical uncertainties,
although we emphasize that the scale-dependence check can only give a lower bound
on the theoretical error.

In Figure \ref{fig:vev}, we perform the inverse of the preceding analysis. This
time, we take $m^2(Q_0)$ as an input given by eq.~(\ref{eq:m2Qreference})
and determine 
$v(Q)$ as an output. Of course, at $Q=Q_0$,
the result is exactly as given in eq.~(\ref{eq:referencemodelMSbar}).
At other $Q$, we obtain $v(Q)_{\rm min}$ by first running all of the other
\MSbar quantities from $Q_0$ to $Q$ and then apply 
eq.~(\ref{eq:Veffmincon}) again.
The results are shown in the left panel of Figure \ref{fig:vev}.
We also obtain $v(Q)_{\rm run}$ by directly running it 
using its RG equations
from $Q_0$. The ratio $v(Q)_{\rm min}/v(Q)_{\rm run}$ is shown in the right panel 
of Figure \ref{fig:vev}. Again, in the best available 
approximation, the scale dependence of the ratio is much smaller
than $10^{-4}$ over the entire range.

\section{The Fermi decay constant\label{sec:GFermi}}
\setcounter{equation}{0}
\setcounter{figure}{0}
\setcounter{table}{0}
\setcounter{footnote}{1}

The Fermi weak decay constant is closely related to the vacuum expectation value,
with $G_F = 1/\sqrt{2} v^2$ at tree-level. Including radiative corrections,
one can write:
\beq
G_F &=& 
\frac{1 + \Delta \overline r}{\sqrt{2} v_{\rm tree}^2} 
\>=\>
\frac{1 + \Delta \widetilde r}{\sqrt{2} v^2} .
\eeq
Expressions for $\Delta \overline r$ have been given at 2-loop
order in the so-called gauge-less limit 
($g^2, g^{\prime 2} \ll g_3^2, y_t^2, \lambda$) in
ref.~\cite{Kniehl:2014yia} and ref.~\cite{Kniehl:2015nwa}, 
using expansions in terms of \MSbar and on-shell quantities respectively,
but in both cases determined in terms of the tree-level VEV. 
The full 2-loop version of $\Delta \overline r$ is quite lengthy, and to our knowledge 
has not appeared in print, but was obtained and presented
within the public computer code {\tt mr} \cite{Kniehl:2016enc}. 
We have obtained the corresponding complete 2-loop result for $\Delta \widetilde r$ in terms of $v$, 
\beq
\Delta \widetilde r &=& 
\frac{1}{16\pi^2} \Delta \widetilde r^{(1)} + 
\frac{1}{(16\pi^2)^2} \Delta \widetilde r^{(2)} + \ldots .
\eeq
The 1-loop order part is
\beq
\Delta \widetilde r^{(1)} &=&
\frac{3}{4} (g^2 - g^{\prime 2}) [A(Z) - A(W)]/(Z-W)
+ \frac{3}{4} \left [(4 g^2 - 24 \lambda) A(W) -g^2 A(h) \right ]/(h-W)
\nonumber \\ &&
\!\!\!\!\!\!\!\!\!
+ 3 [y_t^2 A(t) - y_b^2 A(b)]/(t - b) + 2 A(\tau)/v^2
- (3 g^2 + g^{\prime 2})/8 + (3 y_t^2 + 3 y_b^2 + y_\tau^2)/2
-\lambda
, 
\eeq
where
\beq
Z &=& (g^2 + g^{\prime 2}) v^2/4, \qquad 
W\>=\> g^2 v^2/4, \qquad
h\>=\> 2 \lambda v^2,  
\\
t &=& y_t^2 v^2/2,\qquad b \>=\> y_b^2 v^2/2,\qquad \tau \>=\> y_\tau^2 v^2/2,
\eeq
are the running \MSbar squared masses, and 
\beq
A(x) = x \lnbar(x) - x
\eeq
with
\beq
\lnbar(x) = \ln(x/Q^2).
\eeq
The 2-loop part is
\beq
\Delta \widetilde r^{(2)} &=&
g_3^2 y_t^2 [8 \zeta_2 - 17 - 16 A(t)/t - 12 A(t)^2/t^2]
+ \Delta \widetilde r^{(2)}_{\small\mbox{non-QCD}}
,
\eeq
where $\Delta \widetilde r^{(2)}_{\small \mbox{non-QCD}}$ is again
rather lengthy, and so is provided in its
complete form as an ancillary file {\tt Deltartilde.txt} 
distributed with this paper, rather than in text form here. 
It has the form:
\beq
\Delta \widetilde r^{(2)}_{\small\mbox{non-QCD}} = \sum_j C_j^{(2)} I_j^{(2)} 
+ \sum_{j\leq k} C_{j,k}^{(1,1)} I_j^{(1)} I_k^{(1)} 
+ \sum_{j} C_{j}^{(1)} I_j^{(1)}
+ C^{(0)} 
\eeq
where the lists of 2-loop and 1-loop basis integrals required are:
\beq
I^{(2)} &=& \{
\zeta_2,\> 
I(h,h,h),\> 
I(h,t,t),\>
I(0,h,t),\>
I(0,h,W),\>
I(0,h,Z),\>
I(0,t,W),\>
\nonumber \\ &&
I(0,t,Z),\> 
I(0,W,Z),\>
I(h,h,W),\> 
I(h,W,W),\> 
I(h,W,Z),\> 
I(h,Z,Z),\> 
\nonumber \\ &&
I(t,t,W),\> 
I(t,t,Z),\> 
I(W,W,W),\> 
I(W,W,Z),\> 
I(W,Z,Z)
\} ,
\\
I^{(1)} &=& \{ A(t),\> A(h),\> A(Z),\> A(W) \},
\eeq
with the 2-loop vacuum integral function $I(x,y,z)$ as defined as in 
previous papers e.g.~\cite{Martin:2003qz,TSIL,Martin:2016bgz},
and the coefficients $C_j^{(2)}$, $C_{j,k}^{(1,1)}$, $C_j^{(1)}$, and $C^{(0)}$ are rational
functions of $t$, $h$, $Z$, $W$, and $v$. (The $v$ dependence is $1/v^4$ in each case.)
The Goldstone boson contributions in $\Delta \widetilde r$ have been resummed, so that,
as explained in refs.~\cite{Martin:2014bca,Martin:2017lqn}, the
Higgs squared mass appearing here
is $h\equiv 2 \lambda v^2$, and not $m^2 + 3 \lambda v^2$.
Also, note that $\Delta \widetilde r^{(1)}$ is well-defined in the formal
limits $W\rightarrow Z$, $W \rightarrow h$, and $b \rightarrow t$, despite 
denominators that vanish in those limits. Furthermore, although
$\Delta \widetilde r^{(2)}$ has several individual 
terms with $\lambda$ in the denominator,
once can check that 
the whole expression for $\Delta \widetilde r$ 
is finite in the limit $\lambda \rightarrow 0$, unlike
$\Delta \overline r$. This illustrates the 
absence of $1/\lambda$ effects in the tadpole-free scheme based on $v$; more generally, the
absence of $1/\lambda$ effects provides useful checks on calculations. We have also checked that
$\Delta \widetilde r^{(2)}$ is well-defined in the formal 
limits where $Z-4t$ and $h-W$ and $W-Z$ and $h-4Z$
and $h-4W$ vanish, despite many of the individual coefficients having denominators containing factors of
these quantities. Furthermore, we have checked that $G_F = (1 + \Delta \widetilde r)/{\sqrt{2} v^2}$
is RG scale invariant through 2-loop order, as required by 
its status as a physical observable.

\begin{figure}[!tb]
\begin{minipage}[]{8.6cm}
  \includegraphics[width=8.5cm,angle=0]{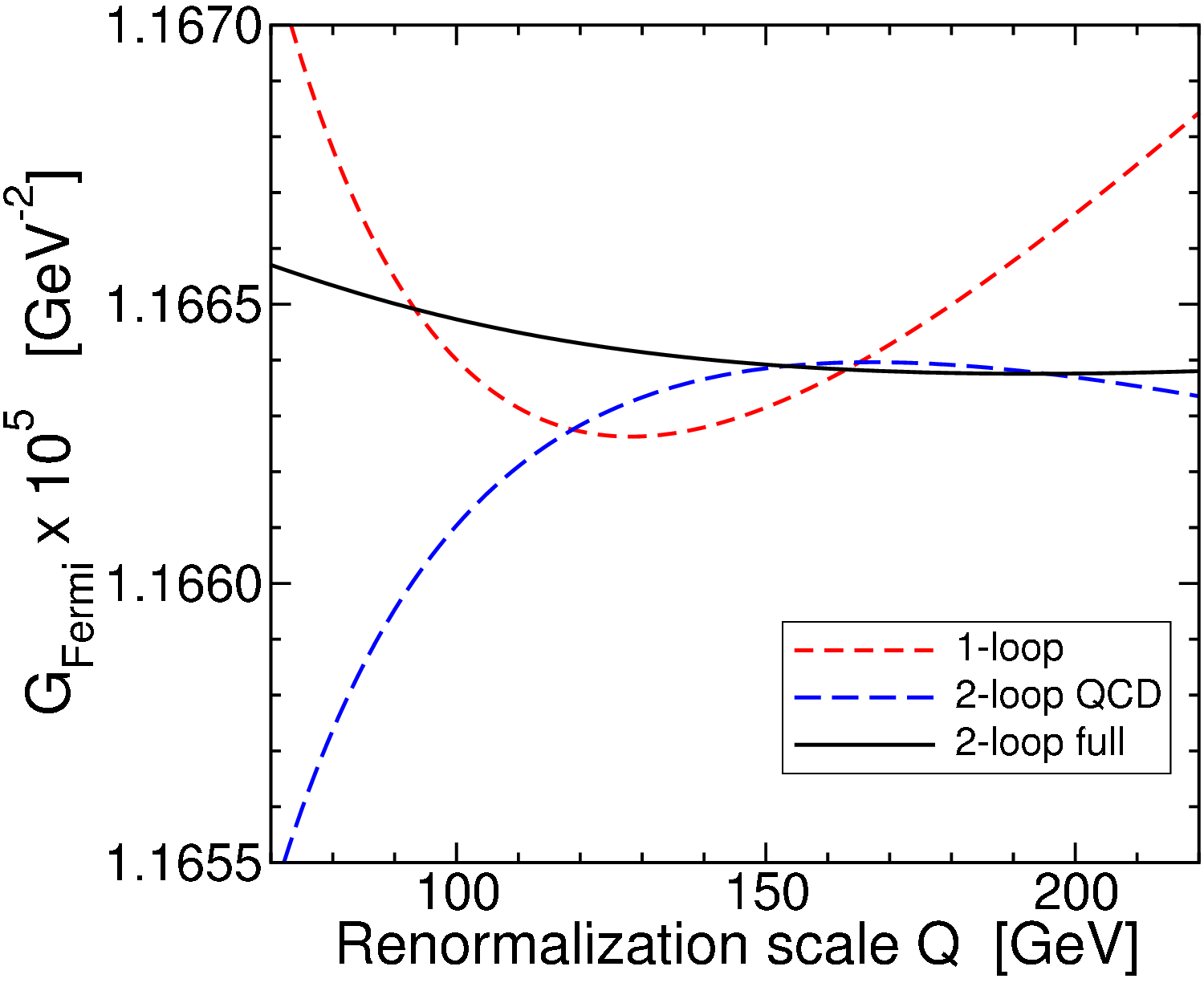}
\end{minipage}
\hspace{0.5cm}
\begin{minipage}[]{6.7cm}
\caption{\label{fig:GFermi} The Fermi constant $G_F$, as a function of
the renormalization scale $Q$ at which it is computed from the \MSbar input parameters, for the reference model point defined at $Q_0 = 173.1$ GeV
in eq.~(\ref{eq:referencemodelMSbar}). 
The short-dashed, long-dashed, and solid lines show 
the results of including the 1-loop, 1-loop plus 2-loop QCD, 
and full 2-loop contributions, respectively.}
\end{minipage}
\end{figure}

This numerical result for $G_F$ in terms of the \MSbar quantities is shown 
in Figure \ref{fig:GFermi} for the benchmark reference model as a function
of the scale $Q$ at which it is computed.
The scale variation is less than 1 part in $10^{-4}$ for $Q$ between 100 and 220 GeV. 
By default, the {\tt SMDR} code evaluates $G_F$ at $Q = M_t$, and so the benchmark point there
agrees exactly with the experimental value. 
The results can also be compared to those of formulas relating $G_F$ to
$M_W$ given by Degrassi, Gambino, and Giardino
in ref.~\cite{Degrassi:2014sxa}, which is larger by a fraction of about 0.0002 (or 0.0001),
provided that $Q$ in our calculation is taken to be close to $M_t$ (or $M_Z$). This corresponds to
a difference in the physical $W$-boson mass of about 
8 MeV (or 4 MeV), less than the current experimental uncertainty in $M_W$.
A further reduction in the purely theoretical sources of uncertainty in our approach could come 
about from including the leading (in $g_3$ and $y_t$) 
3-loop contributions to $G_F$, $M_Z$, and $M_W$. There appear to be no
technical obstacles to performing these calculations; when they become available, they will be included
in the {\tt SMDR} code.

\section{Physical masses of heavy particles\label{sec:MthZW}}
\setcounter{equation}{0}
\setcounter{figure}{0}
\setcounter{table}{0}
\setcounter{footnote}{1}

\begin{figure}[!tb]
  \begin{minipage}[]{0.495\linewidth}
    \includegraphics[width=8.0cm,angle=0]{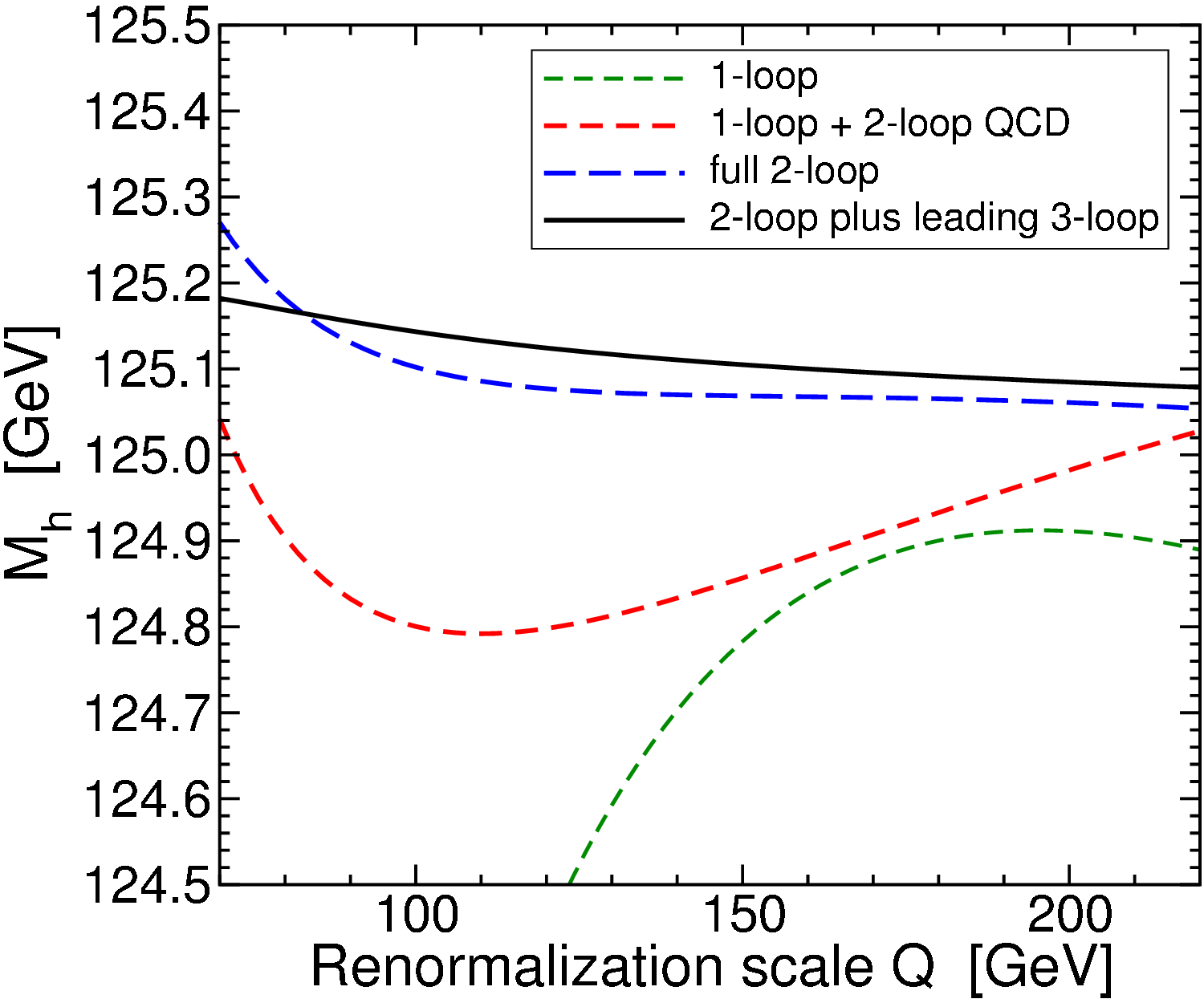}
  \end{minipage}
 \begin{minipage}[]{0.495\linewidth}
    \includegraphics[width=8.0cm,angle=0]{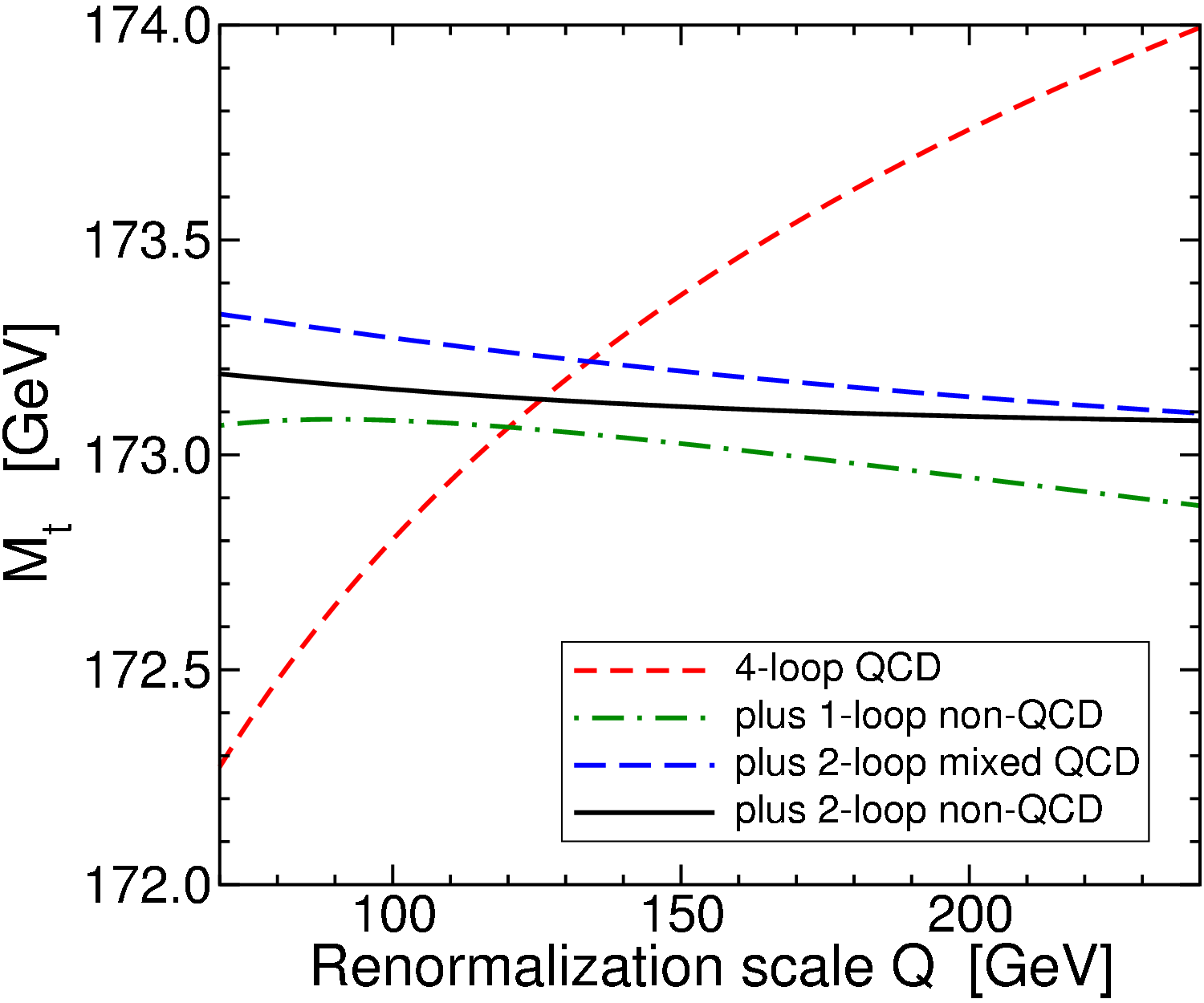}
  \end{minipage}
  \\
  \begin{minipage}[]{0.495\linewidth}
    \includegraphics[width=8.0cm,angle=0]{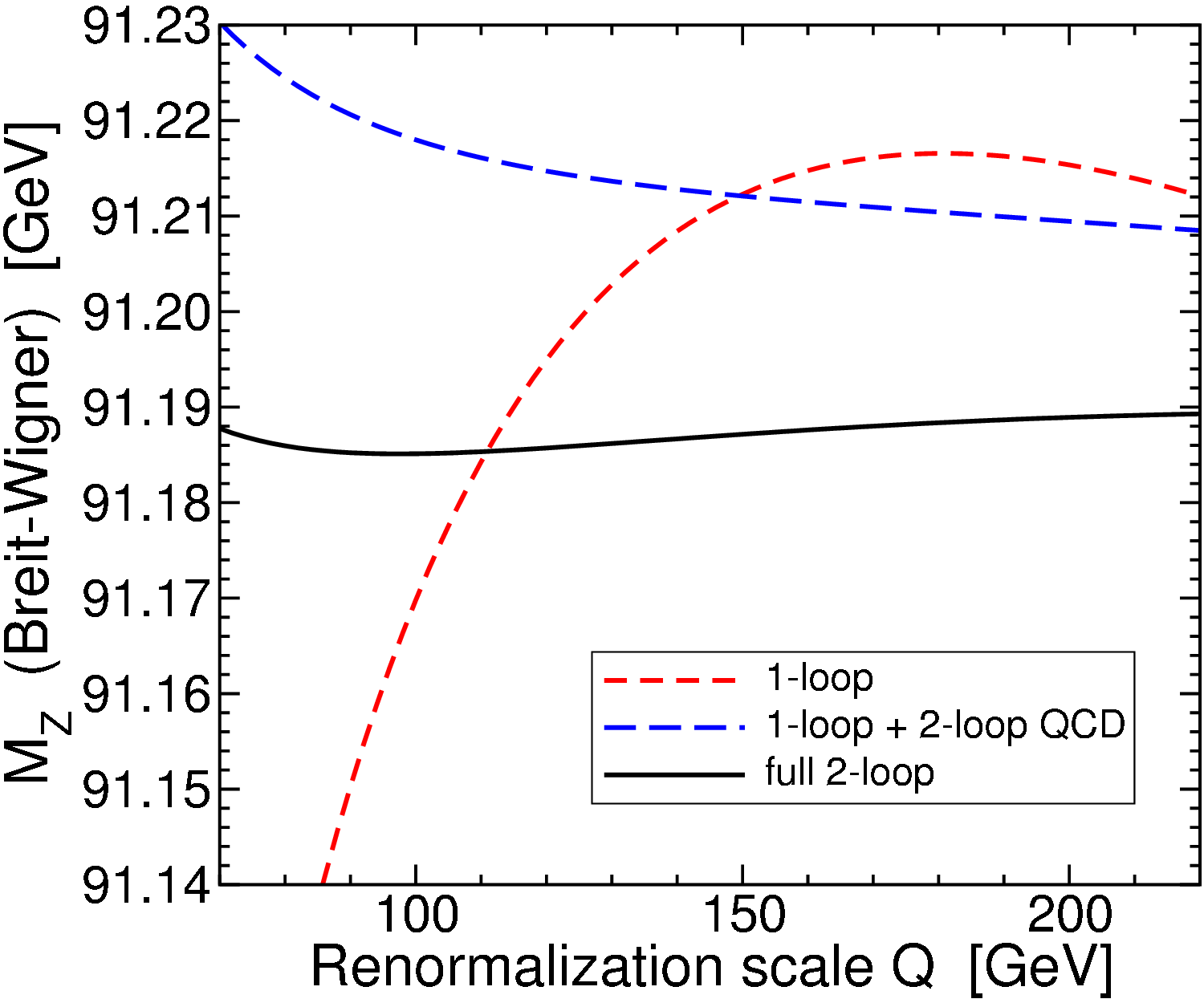}
  \end{minipage}
  \begin{minipage}[]{0.495\linewidth}
    \includegraphics[width=8.0cm,angle=0]{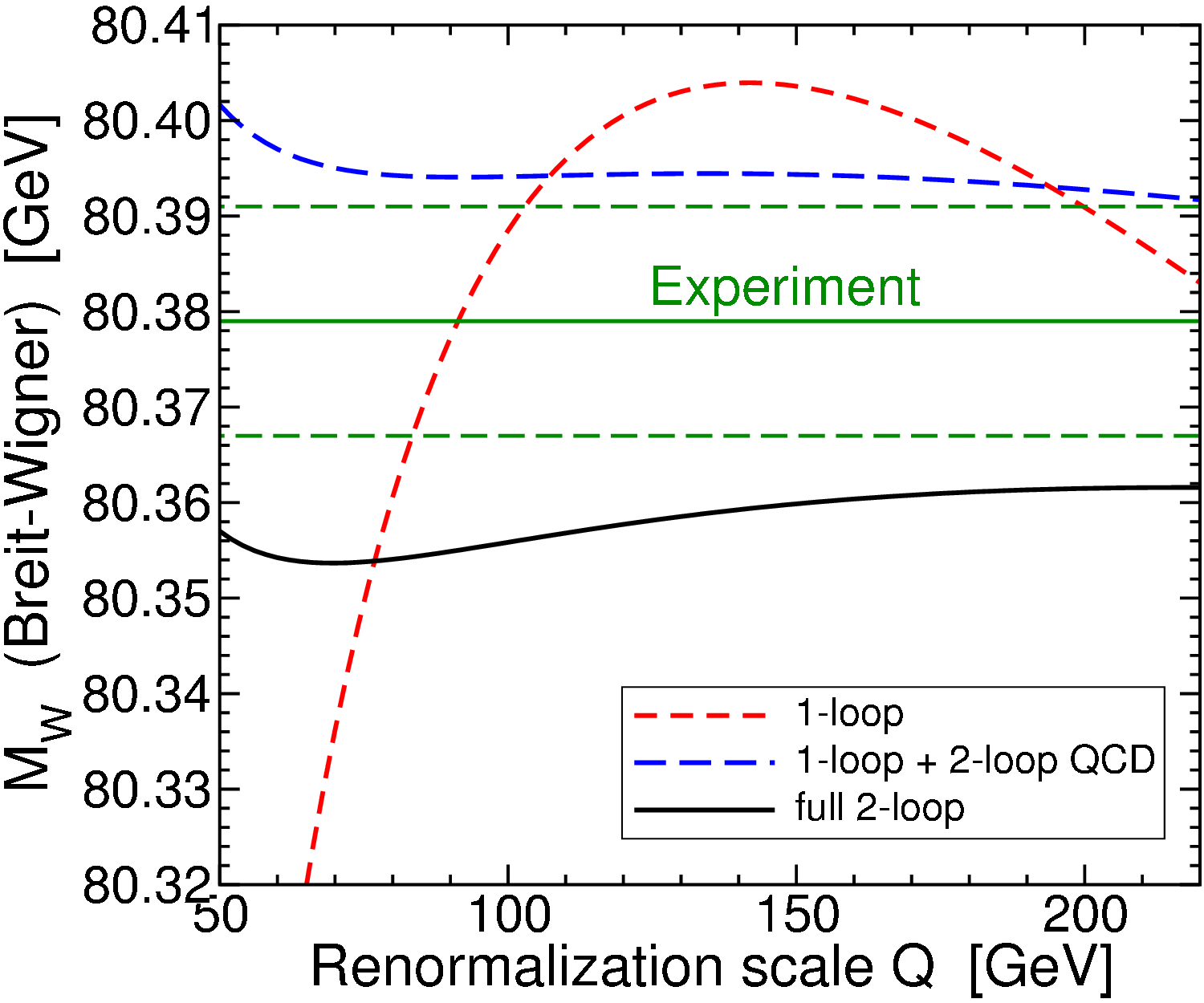}
  \end{minipage}

\begin{center}\begin{minipage}[]{0.95\linewidth}
\caption{\label{fig:MhtZW} Physical masses of the Higgs boson, top quark,
$Z$ boson, and $W$ boson, as functions of the renormalization scale
$Q$ at which they are computed, in various approximations as labeled.
The \MSbar input parameters at $Q$ are determined by RG
evolution from the reference model point
defined at $Q_0 = 173.1$ GeV in eq.~(\ref{eq:referencemodelMSbar}).
In the case of $M_W$, we also show the present experimental central (horizontal solid line) 
and $\pm 1\sigma$ (horizontal dashed lines) values.}
\end{minipage}\end{center}
\end{figure}

For the case of the benchmark
reference model defined in eq.~(\ref{eq:referencemodelMSbar}), 
we show the pole masses of $t$ and $h$
and the Breit-Wigner masses of $W$ and $Z$ in various approximations,
as a function of the renormalization scale $Q$ used for the computation, 
in Figure \ref{fig:MhtZW}.
The results shown are obtained using {\tt SMDR}, which implements
the formulas found in 
refs.~\cite{Martin:2014cxa,Martin:2015lxa,Martin:2015rea,Martin:2016xsp} 
for the tadpole-free pure \MSbar scheme. These papers make use of the
{\tt TSIL} software library in order to numerically evaluate the required
two-loop self-energy basis integrals, using the differential equations method
as described in \cite{Martin:2003qz}, and analytical special cases
found in refs.~\cite{Broadhurst:1987ei, 
Djouadi:1987di, 
Gray:1990yh, 
Scharf:1993ds, 
Berends:1994ed, 
Berends:1997vk, 
Fleischer:1998dw, 
Fleischer:1998nb, 
Davydychev:1998si, 
Jegerlehner:2003py, 
Martin:2003it} and \cite{Martin:2003qz}.

In the case of the Higgs boson pole mass, the $Q$ dependence is seen to be
of order several tens of MeV in Figure \ref{fig:MhtZW}, 
for the best available approximation, which
includes the full 2-loop and leading (in $g_3$ and $y_t$) 3-loop contributions.
However, as we argued in ref.~\cite{Martin:2014cxa}, in the specific case of $M_h$, 
a renormalization scale 
close to $Q=160$ GeV should be made in order to minimize the error from other 3-loop contributions, 
and this choice is used by default in {\tt SMDR}. 

In the case of the top-quark pole mass, in Figure \ref{fig:MhtZW}
we start with the known 4-loop pure QCD
approximation. Although other works often treat the top-quark pole mass using only
QCD effects, the neglect of electroweak corrections is certainly not justified.
Indeed, the 4-loop pure QCD approximation is seen to have a very
large scale dependence of about 1.7 GeV as $Q$ is varied from 70 GeV to 200 GeV.
This shows that failing to include the
electroweak contributions at 1-loop order contributes a very large 
and scale-dependent error, although this is obscured if one also neglects the corresponding
non-QCD contributions in the renormalization group running of the parameters.
Even the 2-loop mixed QCD/electroweak and 
non-QCD effects are roughly of order 200 MeV and 100 MeV, and scale dependent.
By default, the {\tt SMDR} code uses a scale choice $Q = M_t$ when computing $M_t$,
but this can be changed by the user, as for example when making Figure \ref{fig:MhtZW}.

The lower two panels of Figure \ref{fig:MhtZW} show the
dependences of the Breit-Wigner $M_Z$ and $M_W$ on the scale $Q$ at which
they are computed, based on the full 2-loop calculations in 
refs.~\cite{Martin:2015rea,Martin:2015lxa}. The $Q$ dependences are seen to be greatly
reduced by the inclusion of the 2-loop contributions, as expected. The reference
model shown was chosen to reproduce the experimental value of $M_Z$, for
$Q = 160$ GeV. The result for $M_W$ is then a prediction, since it was not
used at all in the determination of the model parameters
in eq.~(\ref{eq:referencemodelMSbar}). Note that
the range of values obtained in Figure \ref{fig:MhtZW} is lower than the
current world average from the Review of Particle Properties in ref.~\cite{RPP}, 
which is $M_W = 80.379 \pm 0.012$ GeV.
This reflects the well-known observation that 
the predicted central value of $M_W$ in the Standard Model is somewhat lower than the observed range, 
but not by enough to draw any firm conclusions about
the validity of the minimal Standard Model. (There is
a long history of calculation of higher-loop contributions 
\cite{vanderBij:1986hy,Djouadi:1987gn,Djouadi:1987di,Kniehl:1989yc,Halzen:1990je,Barbieri:1992nz,Djouadi:1993ss,Fleischer:1993ub,Avdeev:1994db,Chetyrkin:1995ix,Chetyrkin:1995js,Degrassi:1996mg,Freitas:2000gg,vanderBij:2000cg,Freitas:2002ja,Awramik:2002wn,Onishchenko:2002ve,Faisst:2003px,
Awramik:2003ee,Awramik:2003rn,Schroder:2005db,Chetyrkin:2006bj,Boughezal:2006xk}
to the $\rho$ parameter, which gives the $W$ boson mass in terms of the $Z$ boson mass and other on-shell
parameters.)
By default, {\tt SMDR} uses a choice 
$Q = 160$ GeV when computing both 
the $Z$ and $W$ physical masses, but these choices 
can again be modified independently by the
user at run time, as of course was done when making Figure \ref{fig:MhtZW}.
 
\begin{figure}[!b]
  \begin{minipage}[]{0.495\linewidth}
    \includegraphics[width=8.0cm,angle=0]{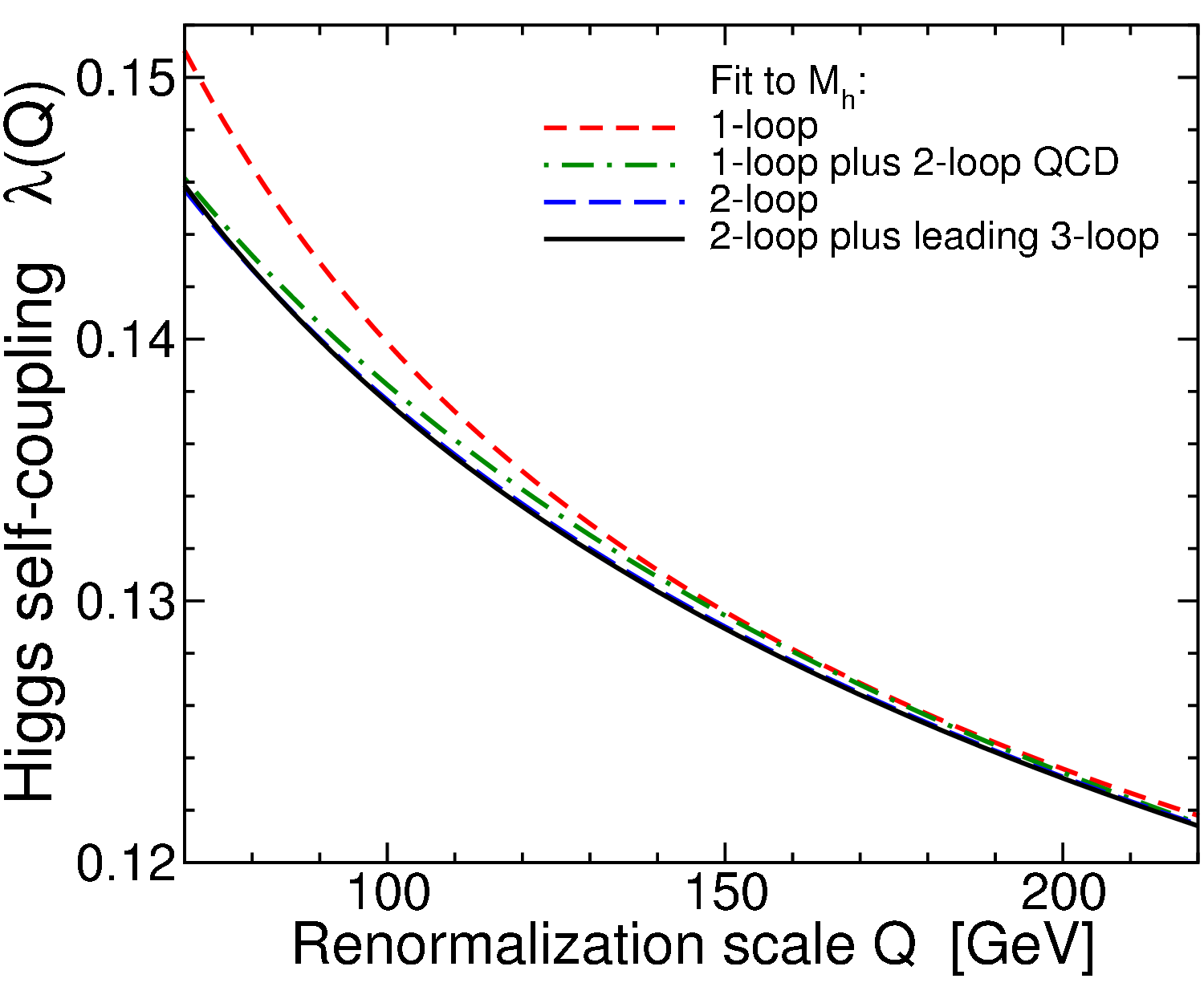}
  \end{minipage}
 \begin{minipage}[]{0.495\linewidth}
    \includegraphics[width=8.0cm,angle=0]{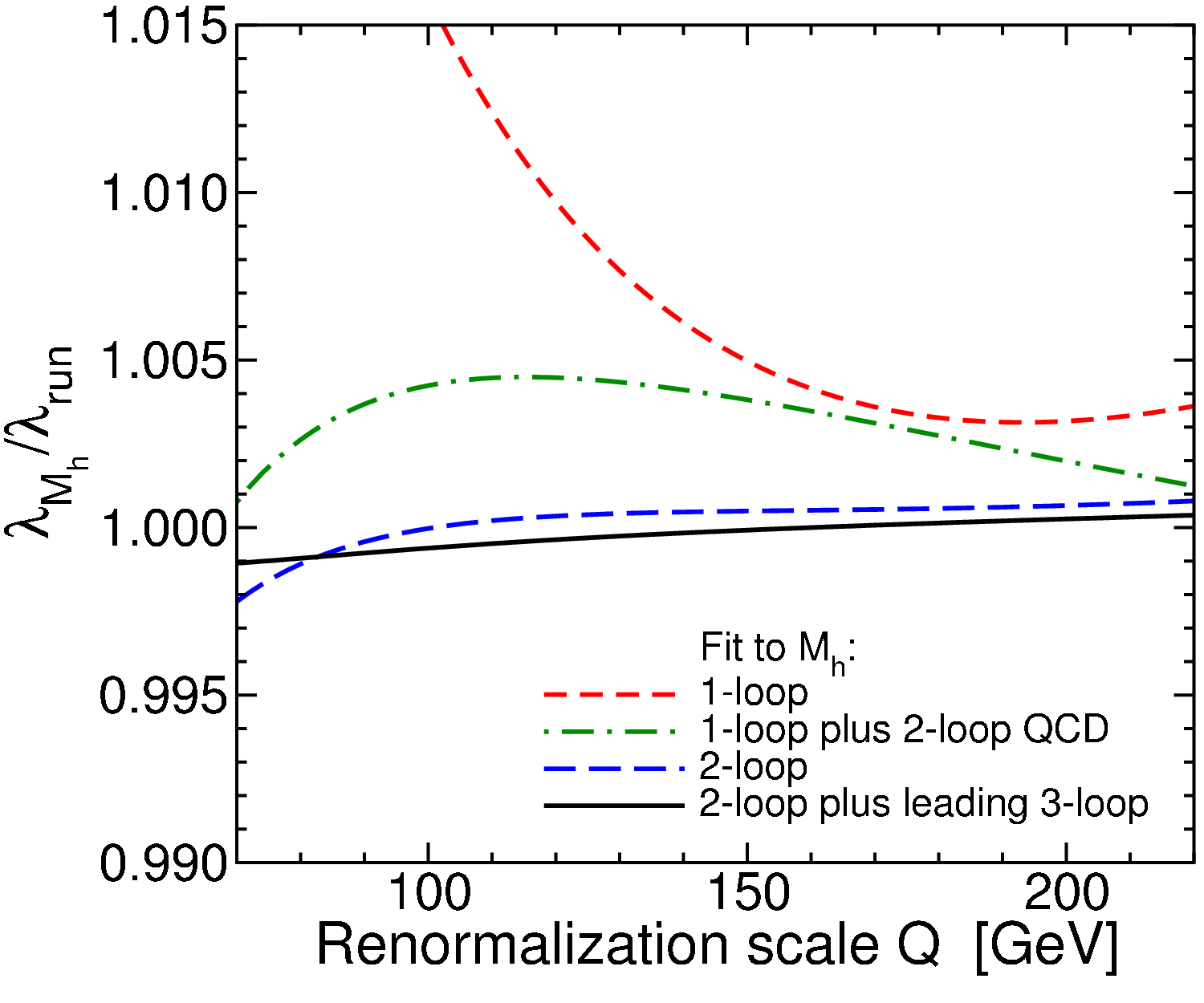}
  \end{minipage}
  
\vspace{-0.5cm}
  
\begin{center}\begin{minipage}[]{0.95\linewidth}
\caption{\label{fig:lambdalowQ} 
The \MSbar Higgs self-coupling $\lambda$, 
as a function of the renormalization scale $Q$, 
for the reference model point
defined at $Q_0 = 173.1$ GeV
in eq.~(\ref{eq:referencemodelMSbar}). 
The other input parameters are obtained from 
the reference model by evolving them
using their RG equations to the scale $Q$, 
where $\lambda(Q)$ is then 
obtained by requiring the Higgs pole mass to be 125.10 GeV.  
In the left panel, results are shown when the calculation of $M_h$ is done 
in the 1-loop, 1-loop plus 2-loop QCD, 
full 2-loop, and 2-loop plus leading 3-loop 
approximations. The right panel shows the results for 
$\lambda(Q)_{\mbox{$M_h$}}/\lambda(Q)_{\mbox{run}}$, 
where $\lambda(Q)_{\mbox{$M_h$}}$ is determined as in the left panel, and
$\lambda(Q)_{\mbox{run}}$ is obtained directly by 
renormalization running from the reference scale $Q_0 = 173.1$ GeV.}
\end{minipage}\end{center}
  \begin{minipage}[]{0.495\linewidth}
    \includegraphics[width=8.0cm,angle=0]{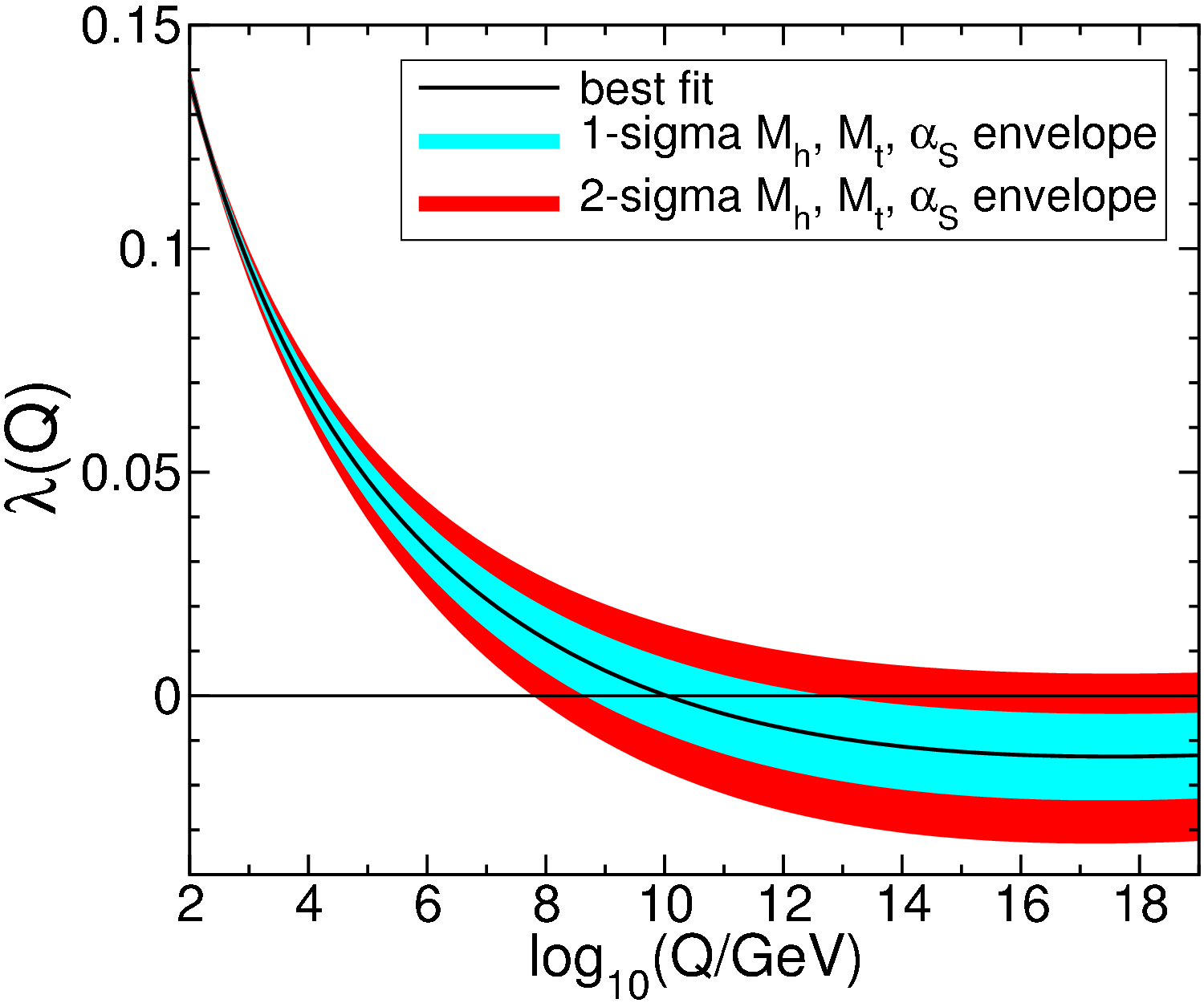}
  \end{minipage}
 \begin{minipage}[]{0.495\linewidth}
    \includegraphics[width=8.0cm,angle=0]{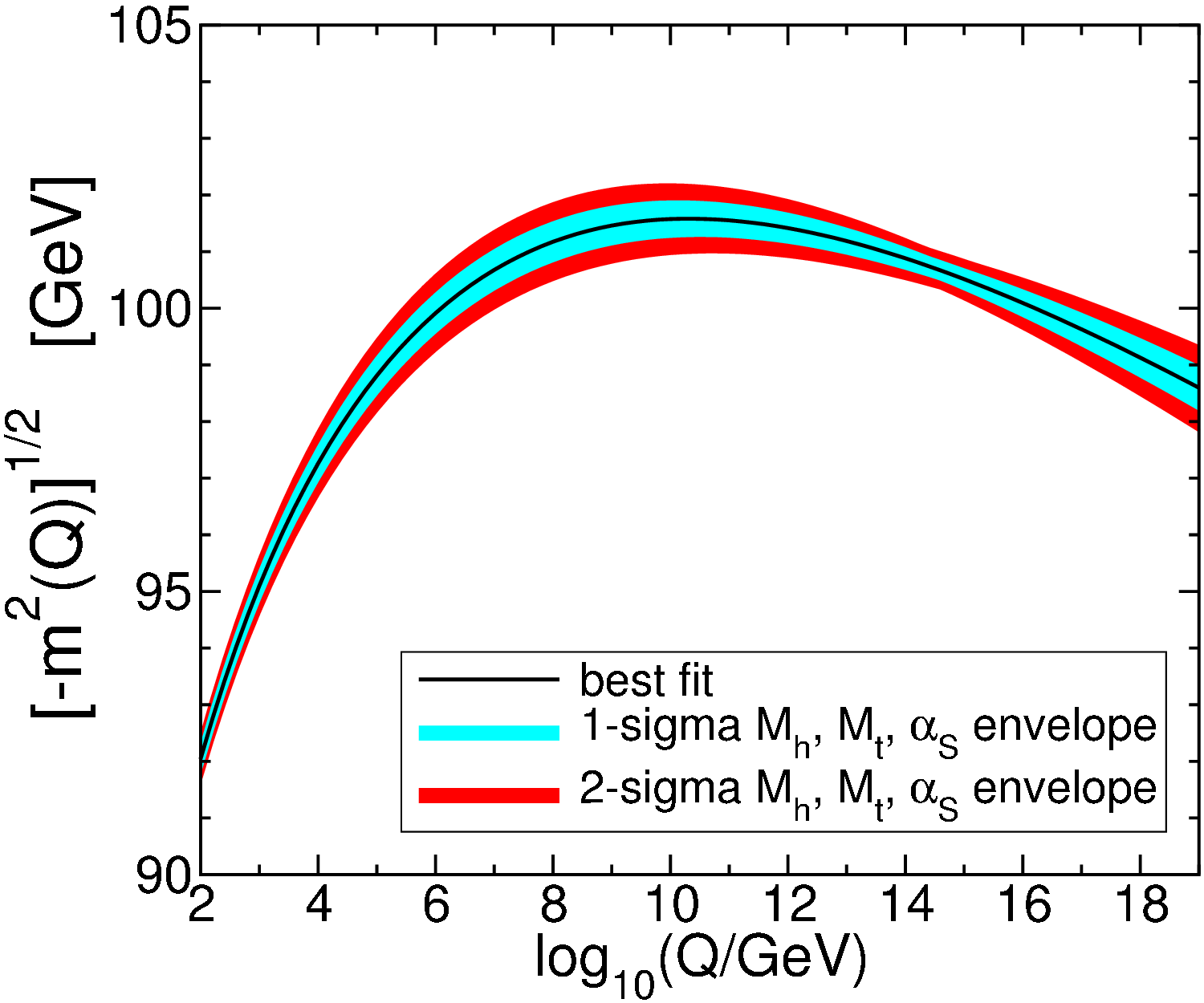}
  \end{minipage}

\vspace{-0.5cm}

\begin{center}\begin{minipage}[]{0.95\linewidth}
\caption{\label{fig:lambdam2highQ}Renormalization group running of the \MSbar 
Higgs potential parameters $\lambda$ and $\sqrt{-m^2}$, as
as a function of the renormalization scale $Q$. The black lines are the central
values obtained from present experimental inputs. Also shown are the envelopes obtained by
varying $M_t$, $M_h$, and $\alpha_S^{(5)}(M_Z)$ within 1-sigma (blue shaded region) 
and 2-sigma (red shaded region)
of their central values. The slight ``pinch" in the envelopes in the right panel 
near $Q = 10^{14}$ GeV is due to a focusing
behavior of the $\alpha_S$ dependence of 
the $m^2(Q)$ renormalization group equation.}
\end{minipage}\end{center}
\end{figure}

The information from the Higgs boson mass $M_h$ can be inverted to obtain the self-coupling $\lambda$,
assuming the minimal Standard Model. This is illustrated in the left panel of 
Figure \ref{fig:lambdalowQ} where we compute
$\lambda(Q)$ at the renormalization scale $Q$ by requiring it to give 
$M_h = 125.10$ GeV, using various approximations
for the calculation of the latter. In the right panel, we then show the ratio of the value $\lambda_{M_h}$ 
obtained in this way to the value $\lambda_{\rm run}$ obtained by RG running it from the value 
in the reference model at $Q_0 = 173.1$ GeV. This ratio is exactly 1 by construction at $Q=Q_0$ in the
approximation used to define the reference model.
In this approximation, the ratio remains less than 1 part in $10^{4}$
over the entire range shown for $Q$.
The parameters $\lambda(Q)$ and $m^2(Q)$ can also be run up to very high scales using the RG equations.
These results are shown in Figure \ref{fig:lambdam2highQ}, including the central value fit as well
as the envelopes resulting from varying each of $M_h$, $M_t$, and $\alpha_S$ 
independently within their 1-sigma and 2-sigma experimentally allowed ranges.
As is now well-known (see for example refs.~\cite{EliasMiro:2011aa} and 
\cite{Bezrukov:2012sa,Degrassi:2012ry,Buttazzo:2013uya}
and references therein), in the best-fit case with $M_h$ near 125 GeV,
$\lambda(Q)$ runs negative at a scale intermediate between the weak scale and the Planck mass,
indicating that our vacuum state may be quasi-stable
if one makes the bold assumption 
that there is really no new physics all the way up to 
mass scales comparable to the scale $Q$ where $\lambda(Q)<0$.

\clearpage

\section{The {\tt SMDR} code\label{sec:code}}
\setcounter{equation}{0}
\setcounter{figure}{0}
\setcounter{table}{0}
\setcounter{footnote}{1}

As noted above, we have
collected our results and methods in the form 
of a public software library written in C, which 
can be used interactively or incorporated into other software, and which is modular 
enough to be easily modified and updated.\footnote{The code {\tt SMDR} subsumes and replaces our
earlier program {\tt SMH}, which evaluated only the Higgs pole mass and 
was described in ref.~\cite{Martin:2014cxa}.} A full description
of how to use {\tt SMDR}, and some example programs, are included with the 
distribution, which is available
for download at \cite{SMDRWWW}. 
For comprehensive information, we refer the reader to 
the file {\tt README.txt}. In this section we give only a brief listing of 
some of the more common user interface variables and functions available. 
Note that these always begin with {\tt SMDR\_} to
avoid naming conflicts with user code.
\begin{itemize}
\item The input values of $Q$ and the \MSbar parameters in eq.~(\ref{eq:MSbarinputs}) 
are specified by global variables {\tt SMDR\_Q\_in},~ {\tt SMDR\_v\_in},~ {\tt SMDR\_lambda\_in},~ 
{\tt SMDR\_g3\_in} etc. 
These can be set or adjusted by the user at any time, but typically remain 
fixed as multiple different tasks are performed, with corresponding temporary global
variables {\tt SMDR\_Q},~ {\tt SMDR\_v},~ {\tt SMDR\_lambda},~ {\tt SMDR\_g3} etc.~used 
for renormalization group running
to various other scales $Q$ and subsequent individual calculations.
\item Renormalization group running in the full, non-decoupled theory is done with the
function {\tt SMDR\_RGeval\_SM()}. In the decoupled QCD+QED theory with 5 quarks and 3 charged leptons,
the evaluation of running parameters (with simultaneous 
decoupling of $t,h,Z,W$ at a scale of choice) is done by
{\tt SMDR\_RGeval\_QCDQED\_53()}. Similarly, evaluation of running parameters at lower scales
including the sequential decoupling of the bottom quark, 
the tau lepton, and the
charm quark, is done by 
{\tt SMDR\_RGeval\_QCDQED\_43()},
{\tt SMDR\_RGeval\_QCDQED\_42()}, and {\tt SMDR\_RGeval\_QCDQED\_32()}, respectively,
where $(5,3)$ and $(4,3)$ and $(4,2)$ and $(3,2)$ refer to the numbers of active quarks and leptons.
\item Minimization of the effective potential to find $m^2(Q)$ from $v(Q)$, 
or vice versa, are accomplished
with functions {\tt SMDR\_Eval\_m2()} or {\tt SMDR\_Eval\_vev()}, respectively.
These make use of the
quantity $\Delta =  \sum_n \Delta_n/(16 \pi^2)^n$ appearing in 
eq.~(\ref{eq:Veffmincon}), which can also be computed separately with {\tt SMDR\_Eval\_vevDelta()}. 
\item Evaluation of the complex pole masses of the four heavy particles is done with functions
{\tt SMDR\_Eval\_Mt()}, {\tt SMDR\_Eval\_Mh()}, {\tt SMDR\_Eval\_MZ()}, and {\tt SMDR\_Eval\_MW()}.
The last two functions also evaluate the variable-width Breit-Wigner masses of $Z$ and $W$, which are 
the traditional ways of reporting those
masses. In each case, one can specify the scale $Q$ at which the computation is performed.
\item Evaluation of the Fermi decay constant is done with the function {\tt SMDR\_Eval\_GFermi()},
again with the computation performed at any specified choice of $Q$.
\item The single function {\tt SMDR\_Eval\_Gauge}() simultaneously evaluates
the Sommerfeld fine structure constant $\alpha_0$
and the RPP ``$\overline{\rm MS}$" scheme
(with only the top-quark decoupled) values $\hat\alpha(Q)$ and $\hat s_W^2(Q)$.
\item The light quark \MSbar masses $m_b(m_b)$, $m_c(m_c)$, $m_s(\mbox{2 GeV})$, 
$m_d(\mbox{2 GeV})$, and $m_u(\mbox{2 GeV})$ are evaluated
using {\tt SMDR\_Eval\_mbmb()}, {\tt SMDR\_Eval\_mcmc()}, and {\tt SMDR\_Eval\_mquarks\_2GeV()}.
\item The charged lepton physical masses can be evaluated using {\tt SMDR\_Eval\_Mtau\_pole()}, 
{\tt SMDR\_Eval\_Mmuon\_pole()}, and
{\tt SMDR\_Eval\_Melectron\_pole()}.
\item A function {\tt SMDR\_Fit\_Inputs()} performs a simultaneous fit 
to all of the \MSbar quantities in eq.~(\ref{eq:MSbarinputs}),
for specified values of the on-shell observable quantities 
(except for $M_W$) in eq.~(\ref{eq:onshellinputs}), providing the results at a specified choice of $Q$.
\item Various utility functions exist for reading parameters from and writing to electronic files.
\item
Our programs {\tt TSIL} \cite{TSIL}
for 2-loop self-energy integrals and {\tt 3VIL} \cite{Martin:2016bgz} for 3-loop vacuum integrals 
are included within the {\tt SMDR} distribution, and so need not
be downloaded separately.
\item
Interfaces for calling {\tt SMDR} from external C or C++ code are included.
\item A command-line program {\tt calc\_all} takes the \MSbar inputs of eq.~(\ref{eq:MSbarinputs}) and outputs all of the
on-shell observables of eq.~(\ref{eq:onshellinputs}).
\item Another command-line program {\tt calc\_fit}  takes the on-shell observables of eq.~(\ref{eq:onshellinputs}) as inputs,
and outputs the results of a fit to the \MSbar inputs of eq.~(\ref{eq:MSbarinputs}), by using the function
{\tt SMDR\_Fit\_Inputs()} mentioned above.
This was used to obtain eq.~(\ref{eq:referencemodelMSbar}).
\end{itemize} 
As examples, the short C programs that produced all of the data used in 
the figures in this paper are included
within the {\tt SMDR} distribution. We also include several other command line programs. 
These should serve to illustrate how to incorporate
{\tt SMDR} into new programs.

\section{Outlook\label{sec:Outlook}}
\setcounter{equation}{0}
\setcounter{figure}{0}
\setcounter{table}{0}
\setcounter{footnote}{1}

In this paper, we have studied the map between the \MSbar 
Lagrangian parameters of the Standard Model and
the observables to which they most closely correspond. In doing so, we have assumed that the 
minimal Standard Model is really the correct theory up to some high 
mass scale, so that new physics contributions effectively decouple. 
With the present absence of evidence at the LHC for new physics, this is at least
a tenable hypothesis, and plausibly 
will remain so for quite some time. We therefore suggest that in the future the
Review of Particle Properties should provide the best-fit values of 
the \MSbar Lagrangian parameters
of the Standard Model in the non-decoupled theory, since these 
fundamentally define the best model that we have
to describe particle physics.

Another useful software package with rather similar aims to {\tt SMDR} but a different 
implementation (including expansion around
what we call $v_{\rm tree}$ rather than $v$) is {\tt mr} \cite{Kniehl:2016enc}. 
There is also a very large number of works that test the whole space of 
electroweak precision observables in different ways; for an incomplete
set of recent references and reviews on this approach, see 
refs.~\cite{Zfitter,Erler:2013xha,Ciuchini:2013pca,Wells:2014pga,
Baak:2014ora,Freitas:2016sty,deBlas:2016ojx,
Haller:2018nnx,Erler:2019hds,Freitas:2019bre}. 
We emphasize that our primary goal here,
of obtaining the best fit to the \MSbar Lagrangian parameters, is
different and complementary to that of testing 
the whole space of electroweak precision observables, as we are not considering 
possible non-negligible contributions from physics beyond the Standard Model.
However, one application is to the matching to new physics models (for example, supersymmetry)
characterized by some mass scale
much larger than the electroweak scale. 
This will necessitate a matching between the high energy theory and the Standard Model as an effective
field theory, including with non-renormalizable operators. For a very incomplete sample of
recent works on this subject, see 
refs.~\cite{Buchmuller:1985jz,Han:2004az,Grzadkowski:2010es,Elias-Miro:2013mua,
Pomarol:2013zra,Chen:2013kfa,Elias-Miro:2013eta,
Ellis:2014dva,Falkowski:2014tna,Henning:2014wua,
Wells:2015uba,Drozd:2015rsp,Ellis:2017jns,Zhang:2016pja,Wells:2017vla,Summ:2018oko}.

New theoretical refinements as well as more accurate experimental measurements will certainly come. 
We have therefore chosen a modular framework in which it should be straightforward to incorporate such 
new developments into the {\tt SMDR} code. For example, we have 
avoided using numerical interpolating formulas 
from approximate fits to analytic formulas, instead opting to provide and use analytical calculations 
directly, up 
to the level of loop integrals that must then be evaluated numerically. This of course results
in longer computation times, but is more transparent and easier to update.
Most of the results presented in this paper are based on calculations that have appeared before, 
but we have provided for the first time a study of the impact of the 
3-loop contributions to the effective potential on the relation 
between the loop-corrected VEV and the
other Lagrangian parameters. We have also provided (in section \ref{sec:GFermi} and an ancillary file,
as well as in the {\tt SMDR} code)
the full 2-loop relation between the loop-corrected VEV
and the Fermi constant, as an alternative to the relation between $G_F$ and the tree-level VEV that
was found in refs.~\cite{Kniehl:2014yia,Kniehl:2015nwa,Kniehl:2016enc}. 
It is clear that significant advances will be needed in order to match the accuracy that can be obtained
at proposed future $e^+e^-$ colliders; for a recent review, see ref.~\cite{Freitas:2019bre}.
Future work in the tadpole-free pure \MSbar scheme
will likely include the leading 3-loop corrections to $M_W$, $M_Z$, and $G_F$.
These and $\Delta \alpha^{(5)}_{\rm had}(M_Z)$ and $M_t$ are the present bottlenecks to accuracy.\\

\noindent {\it Acknowledgments:} We thank James Wells for helpful comments. 
This work was supported in part by the National Science 
Foundation grant number PHY-1719273. DGR is supported by a grant from the Ohio Supercomputer Center.\\

\noindent
{\bf Note added, July 2025:} The following enhancements to the {\tt SMDR} code have been made. See the {\tt CHANGELOG.txt} and {\tt README.txt} files distributed with the code for more information.
\begin{itemize}
\item The default benchmark data are updated with each new version to reflect the latest
results published by the Particle Data Group.
\item In v1.01: the 4-loop contributions to the beta functions for the Standard Model gauge couplings have been completed, using the results of 
ref.~\cite{Davies:2019onf}. These are now used by default.
\item In v1.1: the results of ref.~\cite{Martin:2022qiv} have been included, and are now used by default. Specifically, the Higgs boson pole mass has been enhanced to include the momentum-dependent part of the self-energy at three-loop leading order in QCD ($y_t^2 g_3^4 t$), and with an improved scale dependence of the three-loop part proportional to $y_t^6 t$. The $W$ and $Z$ boson pole and Breit-Wigner masses now include the leading 3-loop QCD contributions.
\item In v1.2: the code now reports complex pole masses in terms of $M$ and $\Gamma$ defined by a different parameterization of the pole masses:
\beq
s_{\rm pole} = (M - i \Gamma/2)^2
\eeq
instead of eq.~(\ref{eq:spole}). This is numerically significant for the $W$ and $Z$ masses.
See the discussion in the Introduction of v2 of ref.~\cite{Martin:2022qiv}, specifically in the two paragraphs surrounding eqs.~(1.1)-(1.7).
\item In v1.3, the calculation of the Fermi constant now includes the leading 3-loop contributions,
in the limit $g_3^2, y_t^2, \lambda \gg g^2, g^{\prime 2}$, as obtained in ref.~\cite{Martin:2025cas}.
\end{itemize}



\end{document}